\documentclass{article}
\usepackage{microtype}
\usepackage{graphicx}
\usepackage{subcaption}
\usepackage{booktabs} 
\usepackage{amsthm}
\usepackage{amsmath}
\usepackage{amssymb}
\usepackage{enumitem}
\usepackage{cases}
\usepackage{mathtools}
\theoremstyle{plain}
\newtheorem{theorem}{Theorem}[section]
\newtheorem{proposition}[theorem]{Proposition}
\newtheorem{lemma}[theorem]{Lemma}

\theoremstyle{definition}
\newtheorem{definition}[theorem]{Definition}

\theoremstyle{remark}
\newtheorem{remark}[theorem]{Remark}
\newtheorem{assum}{Assumption}

\usepackage{hyperref}

\usepackage[accepted]{icml2024}

\usepackage[textsize=tiny]{todonotes}

\icmltitlerunning{Majorization of Conditional Variances}

\begin{document}
\newcommand{\pa}[1]{\mbox{Pa}{#1}}
\newcommand{\bX}{\boldsymbol{X}}
\newcommand{\bvarep}{\boldsymbol{\varepsilon}}
\newcommand\bol{\boldsymbol}
\newcommand{\var}{\mbox{var}}
\newcommand{\chol}[1]{\mbox{chol}{#1}}
\newcommand{\diag}[1]{\mbox{diag}{#1}}
\newcommand{\dichol}[3]{\newcommand{#1}{#2{#3}}}
\newcommand{\indep}{\perp \!\!\! \perp}

\twocolumn[
\icmltitle{Generalized Criterion for Identifiability of Additive Noise Models Using Majorization}



\icmlsetsymbol{equal}{*}

\begin{icmlauthorlist}
\icmlauthor{Aramayis Dallakyan}{to}
\icmlauthor{Yang Ni}{tx}
\end{icmlauthorlist}

\icmlaffiliation{to}{StataCorp, College Station, USA}
\icmlaffiliation{tx}{Department of Statistics, Texas A\&M University, College Station, USA}

\icmlcorrespondingauthor{Aramayis Dallakyan}{dallakyan.aramayis@gmail.com}

\icmlkeywords{DAGs, Causal Discovery, Majorization}

\vskip 0.3in
]



\printAffiliationsAndNotice{}  

\begin{abstract}

The discovery of causal relationships from observational data is very challenging. Many recent approaches
rely on complexity 
or uncertainty concepts to impose constraints on probability distributions, aiming to identify specific classes of directed 
acyclic graph (DAG) models. In this paper, we introduce a novel identifiability criterion for DAGs that places constraints on 
the conditional variances of additive noise models. We demonstrate that this criterion extends and generalizes existing 
identifiability criteria in the literature that employ (conditional) variances as measures of uncertainty in (conditional)
distributions.
For linear Structural Equation Models, we present a new algorithm that leverages the concept of weak majorization applied to
the diagonal elements of the Cholesky factor of the covariance matrix to learn a topological ordering of variables. Through 
extensive simulations and the analysis of bank connectivity data, we provide evidence of the effectiveness of our approach
in successfully recovering DAGs. The code for reproducing the results in this paper is available in Supplementary Materials.

\end{abstract}

\section{Introduction} \label{s:intro}
One of the fundamental problems in science is learning causal relations
from observational data. Directed acyclic graphical (DAG) models 
serve as valuable tools for representing conditional independence and causal relations among random variables. 
Nevertheless, learning DAGs solely from the  
observational data proves to be a difficult problem.
This is primarily due to issues related to identifiability and
the space of the potential DAGs growing super-exponential as the number
of nodes increases.

A commonly used strategy for addressing the non-identifiability issue is to impose restrictions on the joint distribution. For instance, 
methods like PC  \citep{spirtesetal2001}, GES \citep{chickering2003} and 
other related methods \citep{zhangspirtes2015, raskuttiuhler2013}
have demonstrated that, 
under certain assumptions such as faithfulness, it is possible to
recover DAG up to the Markov equivalence class. However, 
it is timely to note that the majority of Markov equivalence
classes contain more than one graph, making it difficult to uniquely determine the true underlying graph.

Recent research has addressed the challenge of recovering causal structures by introducing various constraints aimed at capturing the asymmetry between cause and effect. \citet{shimizuetal06a} propose a method for identifying linear non-Gaussian additive noise models (ANM). \citet{hoyeretal2008, zhangaapo2009, petersetal2012} establish conditions for the identifiability of nonlinear ANMs. \citet{parkraskutti2018, parkpark2019} employ higher-order moments of the conditional distribution to demonstrate the identifiability of DAGs. The ideas
behind these approaches can be succinctly summarized using the concept of complexity
or uncertainty \citep{mooijetal2016, glymouretal2019}. In particular, it is generally anticipated that the complexity of a physical process that generates effect from cause
should be lower in some way than the complexity of the backward process.
In other words, the relationship between two variables in a model is asymmetrical
rather than symmetrical \cite{simon1977}.
These ideas has been explored using Kolmogorov complexity in
\citet{janzingscholkopf2010}. Also see \citet{petersetal2017}. The aforementioned methods harness various complexity measures for both marginal and conditional probability distributions to achieve identifiability.

In a similar vein, several studies \citep{petersbuhlmann2013, lohbuhlmann2014, ghoshal2018, chen2019, park2020} have employed (conditional) variance as a complexity measure to establish identifiability results. Notably, \citet{petersbuhlmann2013} demonstrate the identifiability of linear Structural Equation Models (SEMs) with equal variances. \citet{ghoshalhonorio2017,ghoshal2018, chen2019} note that the arrangement of specific conditional variances implies the identifiability of DAGs. Furthermore, \citet{park2020} introduces more comprehensive criteria concerning conditional variances for nonlinear ANMs with unknown heterogeneous error variances.

In this paper, we establish the identifiability of ANMs by
leveraging the concept of majorization (see Section~\ref{s:prelim} for 
definitions). We prove that our approach constitutes a 
generalization of previous identifiability results that employ conditional variance 
as a complexity measure \citep{petersbuhlmann2013, rajaratnam2013, lohbuhlmann2014, ghoshalhonorio2017, ghoshal2018, chen2019, park2020}. Additionally, we introduce a 
novel DAG structure learning algorithm called \textbf{Ma}jorized 
\textbf{Cho}lesky diagonal (MaCho), which utilizes our established criterion to estimate a topological ordering of a DAG.

\section{Preliminaries} \label{s:prelim}
A DAG $G = (V,E)$ consists of a set of nodes $V = \{1,2, \dots,p\}$ and 
a set of directed edges $E \subset V \times V$ with no directed cycles. 
For the quick introduction to Bayesian Networks and related terminology, see 
Appendix~\ref{s:bn} and \citet{koller2009}. Throughout the paper we assume causal sufficiency, i.e. no
hidden confounders and causal minimality. 

ANMs are a special case of DAG models in which,
for the mean zero random vector $\bX \in R^p$,
the joint distribution is defined by the following structural equation:

\begin{equation} \label{e:anms}
X_j = f_j(X_{\mbox{pa}{j}}) + \varepsilon_j,
\end{equation}
where $\varepsilon_j$'s are independent for $j\in V$ with possibly different distribution with mean zero and hetereogenous variances and $\mbox{pa}_{j}$ denotes the parents of the variable $j$.
A linear SEM is 
a special case of (\ref{e:anms}), in which $f_j$'s are linear. In vector 
formulation, linear SEM can be written as

\begin{equation} \label{e:sem}
    \bX = B \bX + \bvarep,
\end{equation}
where $B \in R^{p \times p}$ is a weighted adjacency matrix such that $\beta_{ij} \neq 0$
indicates $X_j \rightarrow X_i$. From (\ref{e:sem}),
the distribution of $\bX$ is denoted by
$\bX \sim (\boldsymbol{0}, \Sigma)$, where
\begin{equation} \label{e:covmat}
\Sigma = (I - B)^{-1} \Lambda [(I - B)^{-1}]^{'}.
\end{equation}
and $\Lambda$ is a diagonal matrix that contains error variances.

We say a DAG admits a \textbf{topological ordering} $\rho(\cdot)$, if $\rho(j) < \rho(k)$ then $k$ is not an ancestor of $j$,
 i.e. $k \not \in \mbox{An}(j)$, and a $p \times p$ permutation matrix $P_{\rho}$ can be associated such that
  $ P_{\rho}\bol x = (x_{\rho(1)}, \dots, x_{\rho(p)})$, for $\bol x \in R^p$. The existence of a topological ordering leads to the permutation-similarity of $B$ 
  to a strictly lower triangular matrix $ B_{\rho} =  P_{\rho} BP^{'}_{\rho}$ by permuting rows and columns of $B$, respectively \citep{bollen1989} (see Figure~\ref{fig:exdag} in Appendix~\ref{s:bn} for an illustrative example). Therefore,
  \begin{equation} \label{e:ordcov}
      \Sigma_{\rho} = (I - B_{\rho})^{-1} \Lambda_{\rho} [(I - B_{\rho})^{-1}]^{'} = LL^{'} ,
  \end{equation}
  where  $L =  (I - B_{\rho})^{-1} \Lambda^{1/2}_{\rho}$ is a
  Cholesky factor of $\Sigma_{\rho}$. 

  The following Lemma will be useful in Section~\ref{s:alg}.
  The proof is provided in Appendix~\ref{ap:cholvar} for completeness. Also see \citet{rajaratnam2013}, for a similar result
  for autoregressive processes.

\begin{lemma} \label{l:cholvar}
      Suppose a random vector $\bX$ generated from the linear SEM (\ref{e:sem}) with an
      ordering $\rho$, then for $1 \leq j \leq p$ 
\begin{equation}
\begin{aligned}
    L^2_{j,j} = \var(X_{\rho(j)}| X_{\rho(j-1)},\dots, X_{\rho(1)})  
\end{aligned}
\end{equation}
\end{lemma}

\subsection{Majorization} \label{s:major}
The notion of majorization arises in a variety of branches of mathematics and statistics. Informally, majorization describes the notion that the components
of a vector $\bol x \in R^p$ are less spread out or more nearly equal than the 
components of a vector $\bol y \in R^p$.

Formally, let $\bol x_{[\cdot]}$ denote the vector $x$ of a descending order, then we say 
$\bol x$ is \textbf{majorized} by $\bol y$, denoted $\bol x \prec \bol y$, if  $\sum_{i = 1}^{k}x_{[i]} \leq \sum_{i = 1}^k y_{[i]}$, for $k = 1, \dots, p - 1$, and
$\sum_{i = 1}^p x_{[i]} = \sum_{i = 1}^p y_{[i]}$. If 
$\sum_{i = 1}^p x_{[i]} = \sum_{i = 1}^p y_{[i]}$ does not hold, but 
$\sum_{i = 1}^p x_{[i]} \leq \sum_{i = 1}^p y_{[i]}$, then we say
$\bol y$ \textbf{weakly majorizes} $\bol x$, i.e., $\bol x \prec_{w} \bol y$. (Weak) majorization defines
a preordering on $R^p$.

\begin{definition} \label{d:ttrans}
    A linear transformation is called a $T$-transform, if 
    $$T = \lambda I + (1 - \lambda) Q,$$
    where $0 \leq \lambda \leq 1$ and $Q$ is a permutation matrix that interchanges
    two coordinates.
\end{definition}

That is $T \bol x = (x_1, \dots, x_{j-1}, \lambda x_j + (1- \lambda)x_k, x_{j+1}, \dots,
\lambda x_{k} + (1 - \lambda) x_{j}, x_{k+1}, \dots, x_p)^{'}$.

\begin{theorem} (\citet{hardyetal1952}) \label{t:hardy}
For $\bol x, \bol y \in R^{p}$, the following conditions are equivalent
\begin{enumerate} [noitemsep] \label{l:majoreq}
    \item $\bol x \prec_w \bol y$
    \item $\bol x = P \bol y$ for some double substochastic matrix $P$
    \item $\bol x$ can be derived from $ \bol y$ by successive applications of a finite 
    number of $T$-transformations, $\bol x \leq T_1\dots T_k \bol y$
\end{enumerate}
\end{theorem}

The next theorem will be useful to proving identification results in Section~\ref{s:id}.
\begin{theorem} (\citet[Theorems 3.A.8a and 3.C.2.D]{marshalletal11}) \label{t:schur}
Let $f$ be a real-valued function defined on the $A \in R^p$. Then
$$\bol x \prec_w \bol y \implies f(\bol x) < f(\bol y)$$
if and only if $f$ is strictly increasing, symmetric, and strictly convex in $A$.
\end{theorem}

\section{Identifiability via Majorization} \label{s:id} 

For the rest of the paper, we omit the subscript for a topological ordering of the graph $G$, denoted as $\rho_0$, 
whenever there is no potential for confusion. Without loss of generality, we assume $\rho_0 = \{1, \dots, p\}$ and use 
$\rho$ to represent any other ordering. This notation simplification helps avoid the need for the longer notation
${\rho_0(1), \dots, \rho_0(p)}$.


 As we discussed in
Section~\ref{s:intro}, our focus is on identifiability conditions whereas a complexity
measure, the (conditional) variance, has been employed. 
Recall that the concept of complexity or uncertainty implies that 
it is expected that the physical process that is generated using a topological ordering
of variables should be "simpler" than the process generated using other non-topological orderings.
As a measure of the "simplicity" of the process and  to capture the asymmetry between cause and effect,
we impose a weak majorization assumption.
This means if the vector $\bol x$ contains the conditional variances using a topological ordering 
\begin{equation} \label{e:x}
\bol x = \{\var(X_1),\dots,E[\var(X_p|X_1,\dots,X_{p-1})]\}^{'} 
\end{equation}
and $\bol y_{\rho}$   
 contains conditional variances using any other ordering 
\begin{equation} \label{e:y}
\bol y_{\rho} = \{\var(X_{\rho(1)}), \dots, E[\var(X_{\rho(p)}|X_{\rho(1)},\dots,X_{\rho(p-1)})]\}^{'},    
\end{equation}
 then we anticipate $\bol x$ be less spread out then $\bol y_{\rho}$. Using language
introduced in Section~\ref{s:major}, $\bol x \prec_{w} \bol y_{\rho}$ for all permutations $\rho \in \mathfrak{S}_p$,
where $\mathfrak{S}_p$ denotes the group of all permutations $\bX$.
Next, we state the main assumption of this paper.

\begin{assum} \label{a:majorass}
    Let $\bX$ be generated from an ANM (\ref{e:anms}) with DAG $G$, then assume $\bol x \prec_{w} \bol y_{\rho}$ for all permutations $\rho \in \mathfrak{S}_p$, where $\bol x$ and $\bol y_{\rho}$ are defined in (\ref{e:x}) and (\ref{e:y}), respectively.
\end{assum}

\begin{remark}
The notion of varsortability, introduced by \cite{reisachetal2021}, serves as a measure of agreement
between the order of increasing marginal variance and the causal order. In Appendix~\ref{a:varsort}, we 
demonstrate that Assumption~\ref{a:majorass} holds even when the varsortability is low.
\end{remark}

To provide an intuition how the Assumption~\ref{a:majorass} implies identification, we provide examples on bivariate ANMs generated from the following three DAGs: 

\begin{figure}[ht!]
     \centering
     \begin{subfigure}[b]{0.25\linewidth}
         \includegraphics[width=\textwidth]{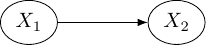}
         \caption{$G_1$}
     \end{subfigure}
     \hfill   
     \begin{subfigure}[b]{0.25\linewidth}
         \includegraphics[width=\textwidth]{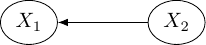}
         \caption{$G_2$}
     \end{subfigure}
     \hfill
        \begin{subfigure}[b]{0.25\linewidth}
         \includegraphics[width=\textwidth]{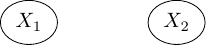}
         \caption{$G_3$}
     \end{subfigure}
        \label{fig:threeDAGs}
\end{figure}

 From Theorem~\ref{t:schur}, the following identification result holds:

\begin{equation} \label{eq:bivid}
\mbox{DAG} := \begin{cases} G_1, & \mbox{if }f(\bol x) < f(\bol y_{\rho})
            \\ G_2, & \mbox{if } f(\bol x) > f(\bol y_{\rho})
            \\ G_3, & \mbox{otherwise}\end{cases},    
\end{equation}

where $\bol x = [\var(X_1), E[\var(X_2|X_1)]]^{'}$, $\bol y_{\rho} = [\var(X_2), E[\var(X_2|X_1)]]^{'}$ and 
in this paper, we use Euclidean norm

\begin{equation} \label{e:maineq}
f(\bol x) = \|\bol x\|_2.    
\end{equation}
For other possible options for $f$, see \citet{marshalletal11}. Empirical results, for bivariate case is provided in Section~\ref{s:numeric}.

The next theorem generalizes the discussed bivariate case. Proof is provided in Appendix~\ref{ap:ident}.
\begin{theorem} \label{t:ident}
    If the Assumption~\ref{a:majorass} holds then DAG $G$
    is uniquely identified.
\end{theorem}


\section{Weak majorization as a generalized condition}

In this section, we show that the weakly majorization Assumption~\ref{a:majorass} is 
a generalization of the existing identification conditions
for nonlinear ANM \citep{park2020}, linear SEM \citep{petersbuhlmann2013, lohbuhlmann2014, ghoshal2018, chen2019, park2020} and autoregressive processes
\citep{rajaratnam2013}.

For the nonlinear ANMs, defined in (\ref{e:anms}), \citet{hoyeretal2008, mooijetal2009, petersetal2012} prove the identifiable classes of ANM by imposing 
restrictions on the functions $f_j$ or the joint distribution. 
\citet{park2020}
used conditional independencies as an uncertainty measure to  propose 
conditions listed in Assumption~\ref{a:ass1}, for identification.
\begin{assum} \label{a:ass1}
Let $\bX$ be generated from an ANM (\ref{e:anms}), then assume for 
$k \in \mbox{De}(j)$ and $l \in \mbox{An}(j)$
\begin{enumerate}
    \item[a.] $\sigma_j^2 < \sigma_k^2 + E(\mbox{var}(E(X_k|X_{\pa{k}})| X_1, \dots, X_{j -1}))$
    \item[b.] $\sigma_j^2 > \sigma_l^2 - E(\mbox{var}(E(X_l|X_1, \dots X_{j-1} \symbol{92} X_l)|X_{\pa{l}}))$
\end{enumerate}
\end{assum}
Moreover, \citet[Theorem 4]{park2020} shows that Assumption~\ref{a:ass1}
is a generalization of conditions proposed in  \citet{petersbuhlmann2013, lohbuhlmann2014, ghoshal2018, chen2019}. Consequently, our goal is to show that
Assumption~\ref{a:ass1} implies the weakly majorization of conditional variances,
given in Assumption~\ref{a:majorass}. 

\begin{theorem} \label{t:main}
    If the conditions of Assumption~\ref{a:ass1} are satisfied, then Assumption~\ref{a:majorass}
    is also satisfied.
\end{theorem}

Proof is provided in Appendix~\ref{ap:main}. 
It is imperative to consider cases where Assumption~\ref{a:ass1} is violated but Assumption~\ref{a:majorass} holds.
 In Appendix~\ref{s:notequiv}, we show that Assumption~\ref{a:ass1}
is a special case of our Assumption~\ref{a:majorass}.
That is there are cases when  Assumption~\ref{a:ass1} is violated but Assumption~\ref{a:majorass} holds.



\section{Algorithm} \label{s:alg}
The proof of Theorem~\ref{t:ident} provides a recipe for learning
topological ordering of a DAG. In particular, we can exploit Theorem~\ref{t:schur}
to compare the estimated value of (\ref{e:maineq}) for all permutation $\rho \in \mathfrak{S}_p$ and choosing the 
one that obtains a minimum value 
$$\min_{\rho}f(y_{\rho}).$$ Unfortunately, for large $p$ this strategy is computationally
infeasible, since $\mathfrak{S}_p$ is large and it requires comparing values of $p!$ permutations. 

In this section, we propose an algorithm, named \textbf{Ma}jorized
\textbf{Cho}lesky Diagonal (MaCho), that learns the topological ordering
of the DAG generated from the linear SEM (\ref{e:sem}) with $O(p^3)$
computational cost.

The intuition behind the proposed algorithm can be explained best by assuming
that the index of the first variable is known, and then considering how the 
algorithm will learn the rest of the ordering on the Cholesky   factor of the population covariance matrix. 
The MaCho algorithm proceeds recursively as follows. Given an already recovered ordering
of the unknown permutation $\rho$, denoted
$\rho_m = \{{\rho(1)},{\rho(2)}, \dots, {\rho(m)}\}$ of size $m < p$, the algorithm chooses the next index
$i_{\rho(m+1)}$ to minimize the $\ell_2$ norm of the diagonal of Cholesky factor
\begin{equation} \label{eq:algf}
\begin{aligned}
f(&\mbox{dichol}[\Sigma_{\rho_m \cup \rho(m+1),\rho_m \cup \rho(m+1)}]) = \\
&\sqrt{L^2_{{\rho(1)},{\rho(1)}} + L^2_{{\rho(2)},{\rho(2)}} + \dots + L^2_{{\rho(m+1)},{\rho(m+1)}}},
\end{aligned}
\end{equation}
where $\mbox{dichol}(A) = \mbox{diag}(\mbox{Cholesky}[A])$ denotes an operator that extracts the diagonal values of the Cholesky factor and $\Sigma_{\rho_m,\rho_m}$ denote the submatrix 
of $\Sigma$ with corresponding indices $\rho_m = \{{\rho(1)},{\rho(2)}, \dots, {\rho(m)}\}$.
Lemma~\ref{l:cholvar} and (\ref{e:maineq}) provide a mathematical justification for (\ref{eq:algf}). 
When the index of the first variable is unknown, in each iteration we check if

\begin{equation} \label{eq:fcomp}
\begin{aligned}
&f(\mbox{dichol}[\Sigma_{\rho_m \cup {\rho(m+1)},\rho_m \cup {\rho(m+1)}}]) \leq \\
&f(\mbox{dichol}[\Sigma_{ {\rho(m+1)} \cup \rho_m,{\rho(m+1)} \cup \rho_m }])
\end{aligned}
\end{equation}

then we add $\rho(m+1)$ at the end of $\rho_m$, otherwise in front of $\rho_m$
and ${\rho(m+1)}$ becomes the first index.
Algorithm~\ref{alg:macho} summarizes the steps.

\begin{algorithm}[tb]
   \caption{MaCho}
   \label{alg:macho}
\begin{algorithmic}
   \STATE {\bfseries Input:} $p \times p$ sample or sparse covariance matrix $S$
   \STATE Select any index as $\rho(1)$.
   \STATE Set $\rho_1 = \{\rho(1)\}$
   \FOR{$i=2$ {\bfseries to} $p$}
   \STATE Choose the next index $$\rho(i) = \min_{i, \dots,p}f(\mbox{dichol}[S_{\rho_{i-1} \cup {\rho(i)},\rho_{i-1} \cup {\rho(i)}}])$$ 
   \IF{(\ref{eq:fcomp}) is TRUE}
   \STATE $\rho_{i} = \{\rho_{i-1}, \rho(i)\}$
   \ELSE
   \STATE $\rho_{i} = \{\rho(i), \rho_{i-1}\}$
   \ENDIF
   \ENDFOR
   \STATE {\bfseries Output: }$\rho_p$
\end{algorithmic}
\end{algorithm}

In practice, when  the number of observations is greater than the dimension of variables, i.e. $n > p$, we suggest to recover  the variable ordering using a modified
version of the Cholesky factor of the sample covariance matrix $S = \hat L \hat L^{'}$, which is
asymptotically unbiased estimator of the population Cholesky factor. The modified version,
is defined as $\tilde L = \hat L \Omega$, where
$$\Omega = \mbox{diag}(1/n, \dots, 1/n-j+1, \dots, 1/ n-p+1).$$
The proof of the next lemma is relegated to Appendix~\ref{ap:statprop}.

\begin{lemma} \label{l:statprop}
Let $\bol X \sim N_p(0, \Sigma)$.
        As $n \rightarrow \infty$, $E(\tilde L) = L$ with probability tending to 1.
\end{lemma}

For high dimensional dataset,
when $n <p$, the covariance matrix can be estimated using a well known sparse
covariance matrix estimation methods such as \citet{bientibshirani2011, wang2014}.
Throughout the paper, when $n < p$, we adopt  covariance graphical lasso algorithm
proposed in \citet{wang2014} and implemented in R package \texttt{covglasso} available via CRAN.

\subsection{Computational details} \label{s:comp}
One of the main computational cost in Algorithm~\ref{alg:macho} is in the computation of
$ \min_{i, \dots,p}f(\mbox{dichol}[S_{\rho_{i-1} \cup {\rho(i)},\rho_{i-1} \cup {\rho(i)}}])$
in each iteration $i$. Fortunately, there is no need to implement Cholesky factorization on the $i \times i$ matrix
in each iteration $i$. 
The new Cholesky factor can be found utilizing the estimated  $(i-1) \times (i-1)$ Cholesky factor $L^{(i-1)}$ from
iteration $i-1$. In particular, the new added row of $L^{(i)}$ can be computed \cite{golub1996}, for $j = 1$
$$L^{(i)}_{i,j} = \frac{S_{i,j}}{L^{(i-1)}_{j,j}},$$

for $j = 2,\dots, i-1$
$$L^{(i)}_{i,j} = \frac{1}{L^{(i-1)}} (S_{i,j} - \sum_{k=1}^{j-1} L^{(i)}_{i,k} * L^{(i-1)}_{j,k})$$
and for $j = i$
$$L_{i,i} = \sqrt{S_{i,i} - \sum_{k=1}^{i-1}(L^{(i)})^2_{i,k}}$$
This requires $O(i^2)$ flops in each iteration $i$. 
The next computational burden is in computing (\ref{eq:fcomp}). Again this can be computed in $O(i^2)$ flops, using the Cholesky factorization
update algorithm in \citet{daviswilliam2005} and \citet[Chapter 12.5]{golub1996}.
From the above discussion, the following lemma easily follows.

\begin{lemma} \label{l:comptime}
    The Algorithm~\ref{alg:macho} has a computational cost $O(p^3)$.
\end{lemma}

\subsection{Estimation algorithm}
Algorithm~\ref{alg:macho} provides a procedure to estimate a topological ordering of the DAG. Given the ordering,
there exist a rich literature on estimating the DAG structure, such as \citet{shojaie2010, khare2016, park2020}. 
Motivated by the encouraging results of thresholding
estimators for the (inverse) covariance matrix \cite{Caietal2016,wangetal2022}, we propose to use
the thresholded version of Convex Sparse Cholesky Selection (CSCS) algorithm introduced in
\citet{khare2016}. CSCS minimizes the following 
objective function

\begin{equation} \label{eq:objfunc}
\mbox{tr}(LL^{'}S) - 2\log|L| + \lambda \sum_{1 \leq j <i \leq p} |L_{ij}|   
\end{equation}

Then our final estimator is 
 \begin{equation} \label{eq:thresh}
    \hat L^{\tau} = \hat L 1(|\hat L| > \tau),
 \end{equation}
 where $1(\cdot)$ is the indicator function, and $\hat L$ is 
 a CSCS estimator.
The next theorem shows the sign consistency of thresholded CSCS under 
the following conditions:

\begin{itemize}
\item{A1} \textit{Marginal sub-Gaussian assumption:} The sample matrix $X \in \mathcal{R}^{n \times p}$ has $n$ independent 
rows with each row drawn from the distribution of a zero-mean random vector $X = (X_1, \dots, X_p)^t$ with covariance $\Sigma$ and sub-Gaussian marginals; i.e.,
\[E[\mbox{exp}(tX_j/ \sqrt{\Sigma_{jj}})] \leq \mbox{exp}(Ct^2) \]
for all $j=1,\dots,p, \, t \leq 0$ and for some constant $C > 0$.
\item{A2} \textit{Sparsity Assumption:} The true Cholesky factor $L \in \mathcal{R}^{p \times p}$ is the lower triangular matrix with 
positive diagonal elements and support $\mathcal{S}(L) =\{(i,j), i \neq j| L_{ij} \neq 0\} $. We denote by $s = |S|$ cardinality of the set $S$.
\item{A3} \textit{Bounded eigenvalues:} There exist a constant $\kappa$ such that
\[0 < \kappa^{-1} \leq \lambda_{min}(L) \leq \lambda_{max}(L) \leq \kappa \]
\item{A4} The minimum edge strength:
\begin{equation}
    \ell_{\min}:= \min_{1 \leq j < i \leq p}|L_{ij}| > c_1 \sqrt{\frac{s\log p}{n}}
\end{equation}
\end{itemize}

\begin{theorem} \label{t:signcons}
    Let the Assumptions 1-3 be satisfied. Further, if Assumption 4
    is satisfied with $c_1 > 2c_2$, where $c_2$ is defined in 
    Lemma~\ref{l:stprop}, the thresholded CSCS estimate $\hat L^{\tau}$
    with threshold level $\tau = c_2 \sqrt{ \frac{s\log p}{n}}$ satisfies:

\begin{equation}
    \mbox{sign}(\hat L^{\tau}) = \mbox{sign}(L_{ij}),\; \forall{i \neq j}
\end{equation}
    with probability tending to 1.
\end{theorem}
The proof is provided in Appendix~\ref{a:cscssign}.
Theorem~\ref{t:signcons} provides an additional support to our usage of thresholded CSCS, since the
estimated model with reversed signs can be misleading and hardly qualifies as a correctly selected model
\cite{pengbin2006}.

\section{Numerical results} \label{s:numeric}
In this section, we measure the performance of the proposed criterion and the algorithm 
using various simulation results. 

\subsection{Bivariate non-linear ANM}

We start our analysis by considering the bivariate case, as discussed in (\ref{eq:bivid}), while incorporating heterogeneous error 
variances using both Gaussian and non-Gaussian distributions. Similar to \citet{park2020}, as a non-linear ANM, we select a polynomial SEM where each
variable is modeled as a 5th-degree polynomial in relation to its parents. In the Gaussian scenario, error variances are randomly chosen 
from the range $\sigma_j^2 \in [0.7, 1.2]$, while in the non-Gaussian case, we employ Gaussian mixture models. For each of these cases, 
we generate 100 sets of samples, each with the number of observations set at $n = 500$.

We then proceed to compare our results with the method proposed in \citet{hoyeretal2008}, referred to as NCD, and utilize the 
\texttt{Causal Discovery} package in Python \citep{kalainathan2019causal}. In our method, the non-linear ANM
is estimated using Random Forest \cite{breiman2001random}. We generate causal effect relationships, such as $x \rightarrow y$, 
$x \leftarrow y$, and no effect, with probabilities of $\{0.4, 0.4, 0.2\}$, respectively. From (\ref{eq:bivid}), 
given that achieving exact equality for the
"no-effect" scenario is challenging, we introduce a tolerance measure denoted as $\mu$, defined as:
$$\mu = \frac{|f(x) - f(y)|}{\max(f(x), f(y))},$$
This measure quantifies the allowed relative difference in magnitude that still qualifies as the absence of a causal effect. 
Based on the simulation results, we recommend $\mu = 0.1$. A similar measure is also defined for the NCD method for the sake of comparison.

Table~\ref{t:bivcase} reports the accuracy of both methods, where accuracy indicates the correct identification of the direction of the causal effect or the identification of a "no-effect" scenario. As evident from the table, our method outperforms NCD for both Gaussian and non-Gaussian
cases.

\begin{table}[ht!]
    \caption{Estimated accuracy for the bivariate case}
    \centering
      \begin{tabular}{c|c|c}
    Method & Gaussian & Non-Gaussian \\
    \hline
       Our  & \textbf{0.95}  & \textbf{0.93} \\ 
        NCD & 0.91 &  0.90 \\
    \end{tabular}
    \label{t:bivcase}
\end{table}

\subsection{Multivariate linear SEM}
For the multivariate linear modeling, we explore three distinct scenarios: 1) Gaussian homogeneity, 
2) Gaussian heterogeneity, and 3) non-Gaussian heterogeneity. In all three scenarios, we generate data based on 
the Structural Equation Model (SEM) described in Equation (\ref{e:sem}).

In the case of Gaussian homogeneity, we set all noise variances equal to $\sigma^2 = 0.7$. For the Gaussian heterogeneity scenario, error variances are randomly selected from the range $\sigma_j^2 \in [0.7, 1.7]$. Finally, in Case 3, data is generated from a mixed Gaussian distribution.

\begin{figure}[ht!]
     \centering
     \begin{subfigure}[b]{0.49\textwidth}
         \centering
         \includegraphics[width=\textwidth, height = 2.5cm]{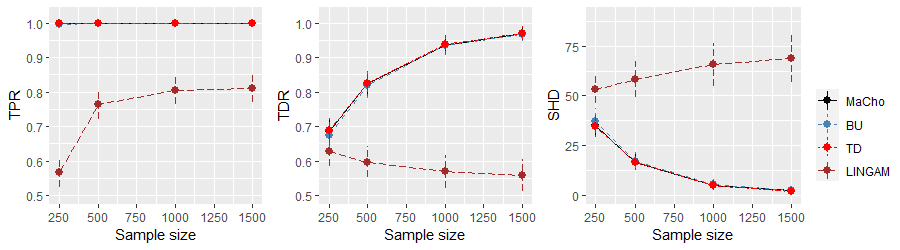}
         \caption{Case 1}
         \label{fig:case1}
     \end{subfigure}
     \vfill
     \begin{subfigure}[b]{0.49\textwidth}
         \centering
         \includegraphics[width=\textwidth, height = 2.5cm]{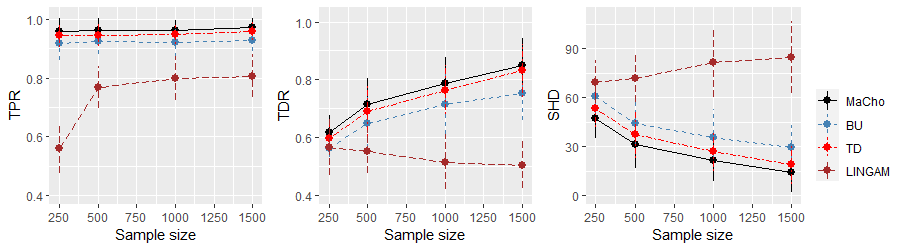}
         \caption{Case 2}
         \label{fig:case2}
     \end{subfigure}
     \vfill
        \begin{subfigure}[b]{0.49\textwidth}
         \centering
         \includegraphics[width=\textwidth, height = 2.5cm]{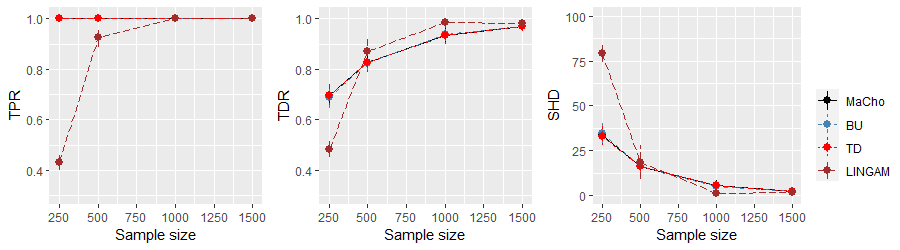}
         \caption{Case 3}
         \label{fig:case3}
     \end{subfigure}
         \caption{Comparison of the proposed algorithm MaCho, BU, TD, and LINGAM algorithms in terms 
         of average TPR, TDR, and SHD for recovering linear SEM with different error variances and $p = 20$.}
        \label{fig:p20result}
\end{figure}

We proceed to compare the performance of our proposed MaCho algorithm with that of the bottom-up (BU) method \cite{ghoshalhonorio2017}, the top-down (TD) method \cite{chen2019}, and the LINGAM algorithm \cite{shimizuetal06a}. For BU and TD algorithms, we employ the \texttt{R} package \texttt{EqVarDAG}, and for LINGAM, we utilize \texttt{pcalg}. 
Both packages are available via CRAN.
For BU and TD, we select the hyperparameter $\lambda$ through a 10-fold cross-validation and 
for MaCho we use procedure described in \citet[Section 3.4]{wangetal2022}.

We evaluate the performance of our algorithm and the comparative methods by varying the sample size $n$ across the set $\{250, 500, 1000, 1500\}$ and the dimensionality $p \in \{20,50, 100\}$. The performance assessment is based on metrics such as the true positive rate (TPR), 
true discovery rate (TDR), and Structural Hamming Distance (SHD), and we conduct this evaluation over 100 replications.


\begin{figure*}[ht]
     \centering
         \includegraphics[width=0.9\textwidth, height = 4.5cm]{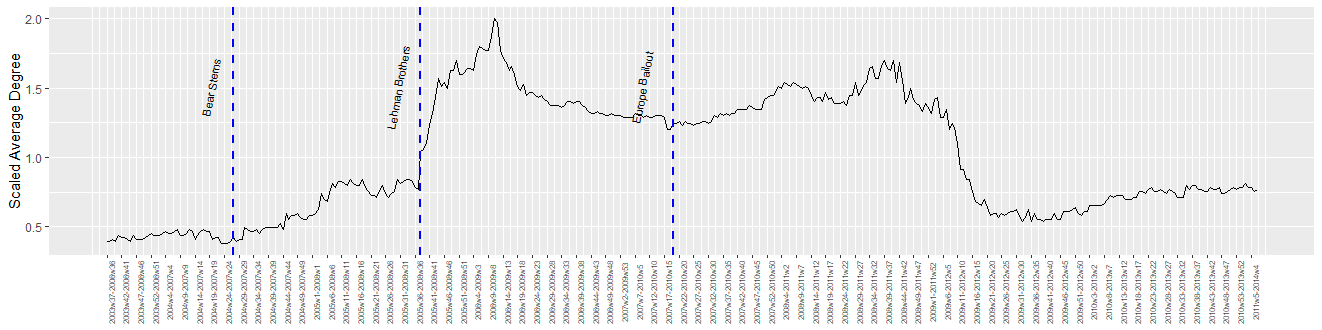}
         \label{fig:y equals x}
        \caption{Evolution of average degree of bank connectedness scaled by their
            historic average (over 2003-2014).}
        \label{fig:avg_degree_macho}
\end{figure*}

Figure~\ref{fig:p20result} presents the results, and it's evident that LINGAM performs the least effectively in both Case 1 and Case 2. In Case 3, LINGAM shows behavior similar to the other algorithms when the sample size n exceeds 1000. Overall, the performance of MaCho, BU,
and TD are quite comparable. For a more detailed illustration of their performance (excluding LINGAM), please refer to Figure~\ref{fig:p20result_LING} in the Appendix. The figure reveals that in Case 2, the performance of MaCho slightly outperforms
BU and TD across all three evaluation metrics.

\begin{figure}[ht]
     \centering
     \begin{subfigure}[b]{0.49\textwidth}
         \centering
         \includegraphics[width=\textwidth, height = 2.5cm]{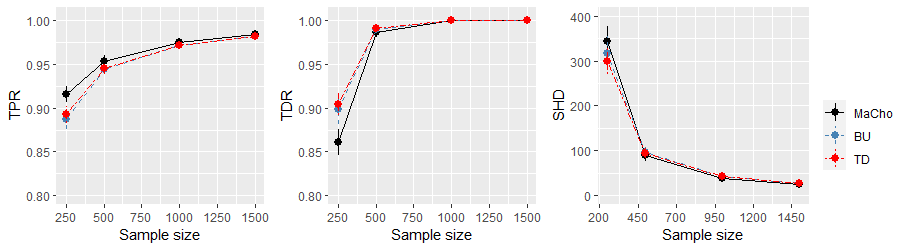}
         \caption{Case 1}
         \label{fig:case1p200}
     \end{subfigure}
     \vfill
     \begin{subfigure}[b]{0.49\textwidth}
         \centering
         \includegraphics[width=\textwidth, height = 2.5cm]{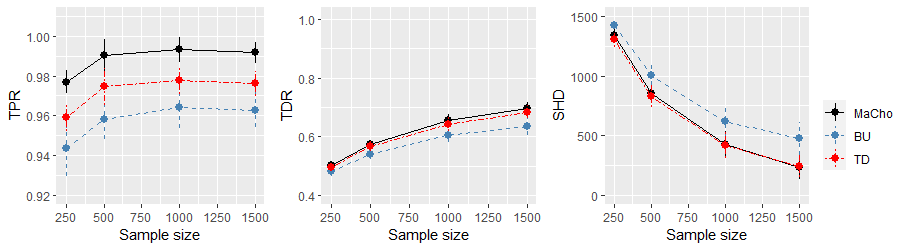}
         \caption{Case 2}
         \label{fig:case2p200}
     \end{subfigure}
     \vfill
        \begin{subfigure}[b]{0.49\textwidth}
         \centering
         \includegraphics[width=\textwidth, height = 2.5cm]{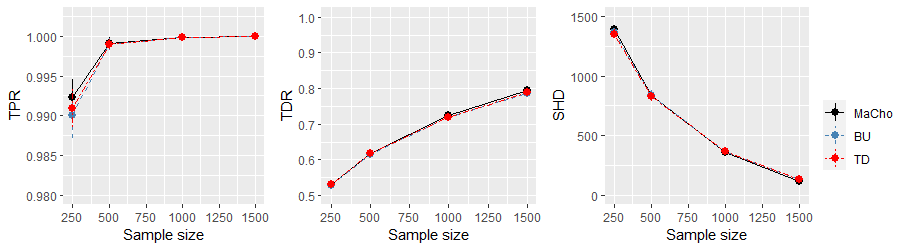}
         \caption{Case 3}
         \label{fig:case3p200}
     \end{subfigure}
         \caption{Comparison of the proposed algorithm MaCho, BU, TD, and LINGAM algorithms in terms 
         of average TPR, TDR, and SHD for recovering linear SEM with different error variances and $p = 200$.}
        \label{fig:p200result}
\end{figure}

In Figure~\ref{fig:p200result}, we present results for $p = 200$ without the LINGAM algorithm.
It is evident that MaCho achieves the highest TPR in all three cases. In terms of the other metrics, 
the performance of MaCho and TD is very close for all cases. Appendix~\ref{a:addsim}
contains results for $p = 50$.

\subsection{Case study: Bank Network Connectedness}
We employ the MaCho algorithm to detect the global bank network connectedness.
The original dataset \citep{demirer2018} comprises 96 banks from 29 developed and
emerging economies (countries),
spanning the period from September 12, 2003, to February 7, 2014. To enhance clarity,
we narrow our focus to economies with more than four banks, resulting in a selection of
54 banks (for further details, please refer to \citet{demirer2018}).
To analyze the data, we adopt a commonly used rolling window  procedure \cite{xueetal2012, basuetal2023}
estimating the model using a 3-year rolling window. 

Our investigation into bank connectedness involves studying the evolution of bank networks
through the average number of direct neighbors.
Figure~\ref{fig:avg_degree_macho} depicts the average degree of the network over 3-year rolling
windows. The plot clearly demonstrates that the connectedness, as measured by the count of neighbors,
increases both before and during significant systematic events. Notably, we observe several prominent
cycles in Figure~\ref{fig:avg_degree_macho}, with three key events marked, the first two corresponding to 
the failures of Bear Stearns and Lehman Brothers, which marked the financial crisis of 2007-2009. 
The final event corresponds to the European Bailout. Our findings align with existing results in the 
literature, including those in \citet[Section 4]{basuetal2023} and \citet{billioetal2012}. 
For comparative analysis, please refer to Appendix~\ref{a:adresult}, where we present results for LINGAM.

\begin{figure*}[ht!]
    \centering
    \includegraphics[width = 0.8\textwidth, height = 7cm]{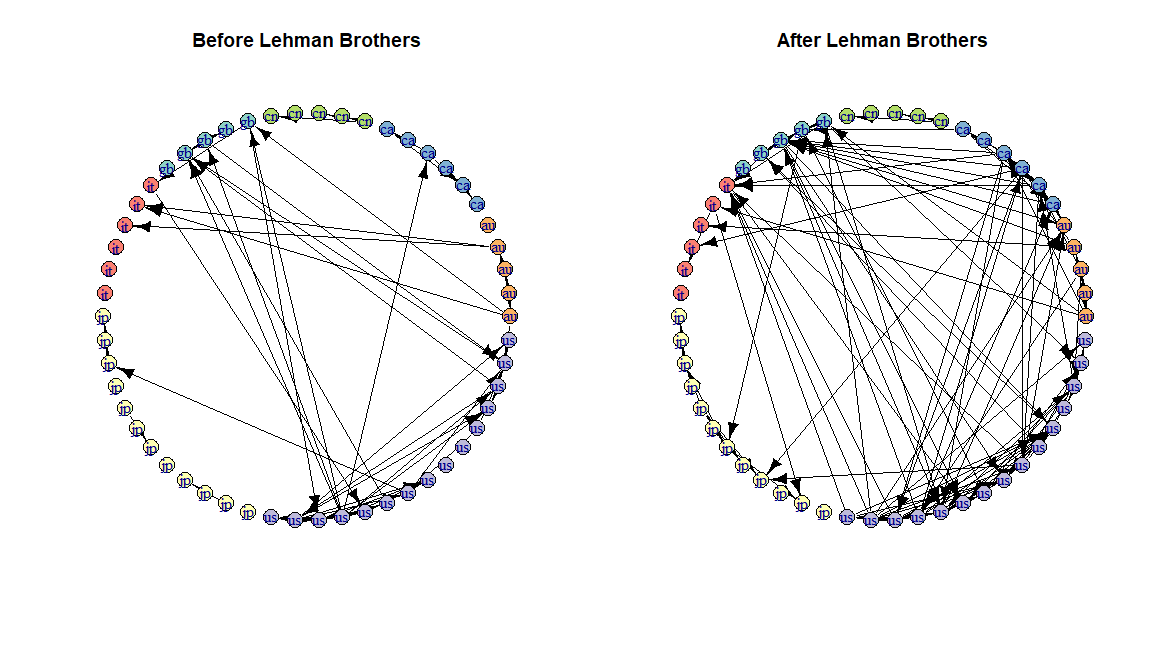}
    \caption{The DAG estimated from MaCho network for the pre and post Lehman Brothers failure.}
    \label{fig:dagmacho}
\end{figure*}

In Figure~\ref{fig:dagmacho}, we present DAGs corresponding to bank connectedness both before and after the Lehman Brothers'
failure. In this figure, the colors of the nodes represent the respective countries to which the banks belong. 
As anticipated, the DAG estimated after the Lehman Brothers' failure exhibits a notably denser structure.

\begin{figure*}[ht!]
     \centering
      \includegraphics[width=0.8\textwidth, height = 7.5cm]{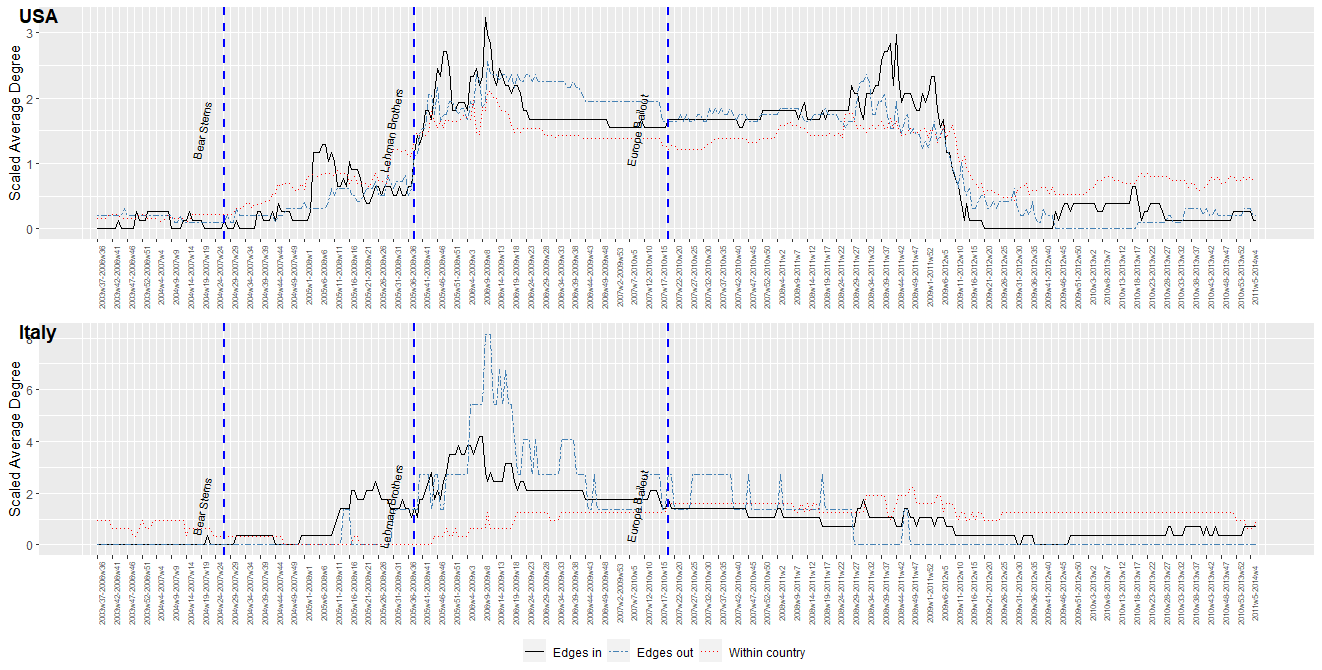}
              \caption{Evolution of average intra and inter-degree of bank connectedness scaled
              by their
            historic average (over 2003- 2014). (Top) USA and (Bottom) Italy.}
        \label{fig:intra}
\end{figure*}

Another intriguing question pertains to how bank connectedness evolved within individual countries.
To address this, we treat each country as a distinct component and estimate both intra- and inter-connectedness.
In Figure~\ref{fig:intra}, the red dotted line represents the average degrees within banks in the USA and Italy.
The black solid and blue dashed lines correspond to the estimated average degrees of in-edges and out-edges of
the country as a component. In this context, in-edges signify changes within the country's causal relationships,
while out-edges signify how banks in the country influence banks in other countries.

A visible contrast emerges in the evolution of network connectedness between these two countries. 
In the USA, all three lines exhibit similar behavior, indicating that connectedness tends to grow before and 
during significant events. However, in Italy, the growth of within connectedness occurs much more gradually 
when compared to the average degree of out-edges.

\section{Conclusion and Discussion}

In this paper, we have introduced a novel identifiability condition for ANMs with heterogeneous error variances.
Our work demonstrates that ANMs can achieve identifiability without the need to assume linearity and Gaussianity. For linear SEMs, we have presented the MaCho algorithm, which is designed for learning the topological ordering of DAGs.
Our findings are supported by various numerical experiments, which empirically confirm that MaCho successfully learns the true topological ordering.

One prominent area for future research involves exploring the testing of the weakly majorization condition for ANMs when the dimensionality $p$ is large. We have demonstrated that empirical testing of the weakly majorization assumption is feasible 
for smaller dimensions, but the computational costs become prohibitively high as $p$ increases.


\bibliography{main}
\bibliographystyle{icml2024}

\newpage
\appendix
\onecolumn

\section{Proof of Lemma~\ref{l:cholvar}} \label{ap:cholvar}
In this section, we show that the diagonal elements of the Cholesky factor of the 
covariance matrix $\Sigma_{\rho}$ have a probabilistic interpretation as the conditional
variances. 

From (\ref{e:sem}), $\bvarep = (I - B_{\rho})^{-1} \bX$. That is for $2 \leq j \leq p$,
$$\varepsilon_{\rho(j)} = X_{\rho(j)} + \sum_{i = 1}^{j-1} U_{\rho(j), \rho_o(i)}X_{\rho(i)},$$
where $U_{\rho_0} = (I - B_{\rho})^{-1}$. Since by definition of SEM, for $\rho(k) < \rho(j)$, 
$\varepsilon_{\rho(j)} \indep X_{\rho(k)}$ then

$$
\begin{aligned}
\var(\varepsilon_{\rho(j)}) &= \var(\varepsilon_{\rho(j)}|X_{\rho(1)}, \dots, {\rho(j-1)}) = \var(X_{\rho(j)} + \sum_{i = 1}^{j-1} U_{\rho(j), \rho(i)}X_{\rho(i)}|X_{\rho(1)}, \dots, X_{\rho(j-1)}) \\
& = \var(X_{\rho(j)}|X_{\rho(1)}, \dots, X_{\rho(j-1)} )
\end{aligned}
$$

Since $\Sigma_{\rho} = L L^{'}$ is the unique Cholesky decomposition, from (\ref{e:ordcov}), the result follows.

\section{Proof of Theorem~\ref{t:ident}} \label{ap:ident}
The first part of the proof adapts a similar idea and technique as Theorem 1 in \citet{petersbuhlmann2013}.
We assume that there are two models as in (\ref{e:anms}) with DAGs $G$ and $G^{'}$ respectively,
such that both induce the same joint distribution. We will show by contradiction that $G = G^{'}$.
By definition, DAGs contain nodes that have no child, such that eliminating such nodes from the graph
leads to a DAG again. Consequently, in both $G$ and $G^{'}$, we remove all nodes that have no children but have the
same parents in both graphs. If there are no remaining nodes, the two graphs are identical and the result follows. 
Otherwise, we obtain two smaller DAGs, which we again call $G$ and $G^{'}$. It is important to note that, there exists
at least one node $L$ in $G$ that has no children and is such that either the parents sets $\mbox{Pa}^{G}_L \neq \mbox{Pa}_L^{G^{'}}$ or $\mbox{Ch}_{L}^{G^{'}} \neq \emptyset$. Therefore, from Markov property of the distribution
with respect to $G$, all other nodes of the graph are independent from $L$ given its parents
$$L \indep \bol X\symbol{92}(\mbox{Pa}_L^{G} \cup L)| \mbox{Pa}_L^{G}$$

\begin{figure}[ht!]
     \centering
     \begin{subfigure}[b]{0.3\linewidth}
         \centering
         \includegraphics[width=0.7\textwidth]{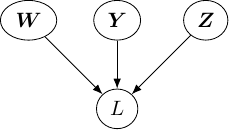}
         \caption{Part of $G$}
     \end{subfigure}
     \hfill   
     \begin{subfigure}[b]{0.3\linewidth}
         \centering
         \includegraphics[width=0.7\textwidth]{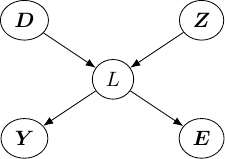}
         \caption{Part of $G^{'}$}
     \end{subfigure}
         \caption{Partition of the nodes in $G$ and $G^{'}$.}
        \label{fig:partition}
\end{figure}

For the rest of the proof, it is informative to partition the parents of $L$ in $G$ into $\bol {Y}, \bol Z$ and $\bol W$
(for illustration, see Figure~\ref{fig:partition}), where
$\bol Z$ are also parents of $L$ in $G^{'}$, the $\bol Y$ are children of $L$ in $G^{'}$, and the $\bol{W}$ are not adjacent to
$L$ in $G^{'}$. Let $\bol D$ be the $G^{'}$ parents of $L$ that are not adjacent to $L$ in $G$ and 
$\bol E$ be the $G^{'}$ children of $L$ that are not adjacent to $L$ in $G$. Thus we have $\mbox{Pa}_L^{G} = \bol{Y} \cup \bol{Z} \cup \bol{W}$
, $\mbox{Ch}_L^{G} = \emptyset$, $\mbox{Pa}_L^{G^{'}} = \bol{Z} \cup \bol{D}$ and $\mbox{Ch}_L^{G^{'}} = \bol{Y} \cup \bol{E}$.

From \citet[Proposition 29]{petersetal2014}, there exist a node $Y \in \bol{Y}$, such that for the sets
$\bol Q := \mbox{Pa}_{L}^{G}\symbol{92}\{Y\},\, \bol{R}:= \mbox{Pa}_{Y}^{G^{'}}\symbol{92}\{L\}$ and $\bol{S}:= \bol{Q} \cup \bol{R}$ we
have $\bol{S} \subset \mbox{Nd}_{L}^G \symbol{92} \{Y\}$ and $\bol{S} \subset \mbox{Nd}_{Y}^{G^{'}}\symbol{92} \{L\}$.
Take any $\bol{s} = (\bol{q,r})$ and denote $L^{*}:= L|_{\bol{S = s}}$ and $Y^{*} := Y|_{\bol{S=s}}$.
From $G$, we can write
$$L^{*} = f_L(\bol{q}, Y^{*}) + \varepsilon_L$$
such that $\varepsilon_L \indep Y^{*}$, since $\bol{S} \subset \mbox{Nd}_{L}^G \symbol{92} \{Y\}$.
Applying the majorization Assumption~\ref{a:majorass} and  denoting $\bol x = [\var(Y^{*}), E[\var(L^{*}|Y^{*})]]^{'}$,
from  Theorem~\ref{t:schur}, $f(\bol x) < f(\tilde {\bol x})$, where $\tilde {\bol x} = [\var(L^{*}), E[\var( Y^{*}| L^{*})]]^{'}$.

Similarly from $G^{'}$ we have
$$Y^{*} = g_{Y}(\bol{r},L^{*}) + \varepsilon_Y$$
such that $\varepsilon_Y \indep L^{*}$, since $\bol{S} \subset \mbox{Nd}_{Y}^{G^{'}}\symbol{92} \{L\}$.
Now, the majorization Assumption~\ref{a:majorass} for $G^{'}$ results in $f(\tilde {\bol x}) < f(\bol x)$,
which is contradiction.

\section{Proof of Theorem~\ref{t:main}} \label{ap:main}
The following results will be used in the proof of the theorem. Recall that, we denoted a topological ordering of the graph $G$
by $\rho_0 = \{1, \dots, p\}$.

\begin{lemma} \label{l:sup1}
    Let $j \in \mbox{De}(i)$, i.e, $i<j$ and Assumption~\ref{a:ass1} is satisfied, then
    \begin{itemize}
        \item[a.] $E[\mbox{var}(X_i|X_1, \dots, X_{i-1})] < E[\mbox{var}(X_j|X_1, \dots, X_{i-1})]$
        \item[b.] $\mbox{var}(X_j|X_1, \dots, X_{j-1}) > \mbox{var}(X_i|X_1, \dots, X_{j}\symbol{92}  X_i)$
        \end{itemize}
\end{lemma}

\paragraph{Proof of a.}$\mbox{}$

From the Law of Total Variance

$$\var(X_j|X_1, \dots, X_{i-1}) = E[\var(X_j|\pa{j})] + \var(E[X_j|\pa{j}]|X_1, \dots, X_{i-1})$$
After taking expectation from both sides and using Assumption~\ref{a:ass1}(a) the result follows.

\paragraph{Proof of b.}$\mbox{}$
From the Law of Total Variance

$$E[\var(X_i|\mbox{pa}_{i})] = \var(X_i|X_1, \dots, X_{j}\symbol{92} X_i) + E[\var(E\{X_i|X_1, \dots, X_{j}\symbol{92} X_i\}|\mbox{pa}{i})]$$

Then, the result directly follows from Assumption~\ref{a:ass1}(b).

\begin{lemma} \label{l:sup2}
    Let $\bX$ be generated from an ANM (\ref{e:anms}) with DAG $G$ and $j \in \mbox{De}(i)$, then
    \begin{itemize}
        \item[a.] $\mbox{var}(X_j|X_1, \dots, X_{i-1}) \geq \mbox{var}(X_j|X_1, \dots, X_{j-1})$
        \item[b.] $\mbox{var}(X_i|X_1, \dots, X_{i-1}) \geq E[\mbox{var}(X_i|X_1, \dots, X_{j}\symbol{92} X_i)]$
    \end{itemize}
    
\end{lemma}

\paragraph{Proof of a.}$\mbox{}$

From the definition of ANM (\ref{e:anms}) for $k < j$, $\varepsilon_j \indep X_k$ and $\var(\varepsilon_j) = \var(X_j|X_1, \dots, X_{j-1})$.
Define $\mbox{pa}{j} = U \cup V$, such that $U \cap V = \emptyset$, $U \subset \{X_1, \dots, X_{j-1}\}$ and for all 
$v \in V$, $v \not \in \{X_1, \dots, X_{i-1}\}$.
Now, from (\ref{e:anms}) and $i < j$

$$\var(X_j|X_1, \dots, X_{i-1}) = \var(f_j(U,V)|X_1, \dots, X_{i-1}) + \var(\varepsilon_j)$$
Since $\var(f_j(U,V)|X_1, \dots, X_{i-1}) \geq 0$, the result follows.

\paragraph{Proof of b.}$\mbox{}$
This is a general result from a well known method of  variance reduction by conditioning and 
follows from the Law of Total Variance.

\paragraph{Proof of Theorem~\ref{t:main}} $\mbox{}$

We show that if the Assumption~\ref{a:ass1} is satisfied then there exist finite number of T-transformations such 
that $\bol{x} \leq T_1\dots T_k \bol{y}$ (see Theorem~\ref{t:hardy}).
As a consequence of \citet[Lemma B.1 and 3.A.5]{marshalletal11}, without loss of generality, to proof results of 
majorization we can consider only cases when $\bol{x}$ and $\bol{y}$ differ in only two components. 

Consider T-transformation $T = \lambda I + (1- \lambda)Q$, where $Q$ is a permutation matrix that interchanges
two coordinates $i$ and $j$, such that $j \in \mbox{De}(i)$ in a topological ordering. That is 
$[X_1,\dots,X_i,\dots, X_j,\dots, X_p]^{'}$ is permuted to $[X_1,\dots,X_j, \dots,X_i, \dots, X_p]^{'}$.
Let $\tilde \bX = Q\bX$.
Using the above notation we define
\begin{equation} \bol x = [\var(X_1), \dots,\var(X_i|\{1,\dots,i-1\}), \dots, \var(X_j|\{1, \dots, j-1\}), 
\dots,\var(X_p|\{1,\dots,p-1\}]^{'}, 
\end{equation}
and 
\begin{equation} \tilde {\bol x} = [\var(X_1), \dots,\var(X_j|\{1,\dots,i-1\}), \dots, \var(X_i|\{1, \dots, j\}\symbol{92} i), 
\dots,\var(X_p|\{1,\dots,p-1\}]^{'}, 
\end{equation}

where we denote $\var(X_i|X_m,X_k)$ as $\var(X_i|\{m,k\})$ and $ \tilde {\bol x}$   
 contains conditional variances using the new ordering with $i$ and $j$ interchanged. 
 From Theorem~\ref{t:hardy} and the definition of T-transformation, our goal is to show that
\begin{equation} \label{eq:proofii}
\bol x_i \leq \lambda \tilde {\bol x_i} + (1 - \lambda) \tilde {\bol x_j}    
\end{equation}
 and 
 \begin{equation} \label{eq:proofjj}
\bol x_j \leq \lambda \tilde {\bol x_j} + (1 - \lambda) \tilde {\bol x_i},    
 \end{equation}
 
where $\bol x_i = \var(X_i|\{1, \dots, i-1\}$, $ \tilde {\bol x_i} = \var(X_j|\{1, \dots, i-1\}$,
$\bol x_j = \var(X_j|\{1, \dots, j-1\})$ and $ \tilde {\bol x_j}= \var(X_i|\{1, \dots,j\symbol{92}i\}$.

From Lemma~\ref{l:sup1} 
\begin{equation} \label{eq:ineq1}
\bol x_i < \tilde {\bol x_i} \; \mbox{and}\; \tilde {\bol x_j} < \bol x_j.    
\end{equation}
Similarly, from 
Lemma~\ref{l:sup2} 
\begin{equation} \label{eq:ineq2}
\bol x_j \leq \tilde {\bol x_i}\; \mbox{and}\; \tilde {\bol x_j} \leq \bol x_i    
\end{equation}
We examine the following cases

\paragraph{Case 1: $\bol x_i \geq \bol x_j$} $\mbox{}$

From (\ref{eq:ineq1}) and (\ref{eq:ineq2})

$$\tilde {\bol x_j} < \bol x_j \leq \bol x_i < \tilde {\bol x_i}$$
Let $\delta = \min(\tilde {\bol x_i} - \bol x_i, \bol x_j - \tilde {\bol x_j})$
and $1 - \lambda = \delta / (\tilde {\bol x_i} - \tilde {\bol x_j})$. Then (\ref{eq:proofii}) and (\ref{eq:proofjj}) follow.

\paragraph{Case 2: $\bol x_i < \bol x_j$} $\mbox{}$

Then 
$\tilde {\bol x_j} \leq \bol x_i < \bol x_j \leq \tilde {\bol x_i}$ and by defining $\delta$ and $\lambda$
accordingly the result follow.

\section{Majorization and \cite{park2020} identifiability conditions}
\label{s:notequiv}
In Theorem~\ref{t:main} we showed that if \cite{park2020} identifiability conditions (Assumption~\ref{a:ass1}) are satisfied
then Assumption~\ref{a:majorass} is also satisfied. In this section, through a toy example we show that those assumptions are not equivalent, i.e.
it is possible that the former is violated but the later is still satisfied.

Consider the following data generation process 

\begin{equation} \label{eq:dgp}
    \begin{aligned}
        X_1 &:= \varepsilon_1\\
        X_2 &:= \beta_{12}X_1 + \varepsilon_2\\
        X_3 &:= \beta_{13}X_1 + \varepsilon_3,     
    \end{aligned}
\end{equation}
where $\sigma^2_1 = 4$, $\sigma^2_2 = 3$, $\sigma^2_3 = 1$, $\beta_{12} = -0.5$
and $\beta_{13} = 0.9$. That is $x = \{4,3,1\}$, where $x$ is defined as in 
(\ref{e:x}).

One of the \citet{park2020} assumptions narrows down to
$$\sigma^2_1 < \sigma_2^2 + \beta_{12}^2 \sigma_1^2$$
However, $4 \not < 3 + 0.25 \times 4$.

To show that the Assumption~\ref{a:majorass} holds, instead of computing the conditional variances 
for all possible permutations and checking the weak majorization, we rely on procedure introduced in Section~\ref{s:quickmajor} to check that the weak majorization
assumption indeed holds for all possible permutation \footnote{In the 
Supplementary Material, we provide the function $\texttt{test\_major()}$, which checks Assumption~\ref{a:majorass} as described in Section~\ref{s:quickmajor}.}.
Table~\ref{t:majorcheck} reports vectors $y_{\rho}$, where for each permutation $\rho \in \mathfrak{S}_p$,
$y_{\rho}$ is defined as in (\ref{eq:esty}). As can be seen, from Proposition~\ref{p:majcheck},
for each $\rho$,  $x \prec_w y_{\rho}$ is true.

\begin{table}[ht!]
    \centering
    \begin{tabular}{c|c}
    $\rho$ & $y_{\rho}$ \\
    \hline
         \{1,3,2\} & \{4,3,1\}  \\
        \{3,1,2\} & \{4.24,3,0.94\}  \\
        \{3,2,1\} & \{4.24,3.23,0.87\}  \\
        \{2,3,1\} & \{4,3.43,0.87\}  \\
        \{2,1,3\} & \{4,3,1\}
    \end{tabular}
    \caption{Estimated vectors $y_{\rho}$ for the toy example.}
    \label{t:majorcheck}
\end{table}

\subsection{Check of majorization assumption for linear ANMs} \label{s:quickmajor}
The following proposition will be useful to check the weak majorization assumption.
Let $\mathcal{D} = \{z: z_{1}\geq \cdots \geq z_n\}$ and
\begin{proposition}(\citet[Proposition B.1.A]{marshalletal11}) \label{p:majcheck}
    Let $x \in \mathcal{D}_{++}$ and $\sum{x_i} \leq \sum{y_i}$, if
    for some $k,\, 1 \leq k \leq n,\, x_{i} \leq y_i$ for $i = 1, \dots,k$ and $x_i \geq y_i$
    for $i = k+1, \dots, n$ then $x \prec_w y$.
\end{proposition}

In this section, we rely on Lemma~\ref{l:cholvar} to propose a quick way to check whether a
majorization is satisfied for the simulated data generated from the linear ANM:

$$\bX = B\bX + \bvarep$$

Let 

\begin{equation} \label{eq:truex}
    x= \mbox{diag(var}(\bvarep))
\end{equation} 

be the diagonal elements of the covariance matrix, and $\Sigma_{\rho}$ is a covariance matrix of $\bX$ after row and column
permutation for some $\rho \in \mathfrak{S}_p$. Define 

\begin{equation} \label{eq:esty}
y = \mbox{diag(chol(}\Sigma_{\rho})^2),    
\end{equation}
 
then, from Lemma~\ref{l:cholvar} the majorization assumption holds if $x \prec_w y$ for all $\rho \in \mathfrak{S}_p$. We summarize the Assumption~\ref{a:majorass} check in 
Algorithm~\ref{alg:ismajor}

\begin{algorithm}[th!] 
   \caption{IsMajorized?}
   \label{alg:ismajor}
\begin{algorithmic}
   \STATE {\bfseries Input:} $x,y$ as in (\ref{eq:truex}) and (\ref{eq:esty})
   \STATE Sort $x$ in a descending order
   \IF{criterion in Proposition~\ref{p:majcheck} holds}
   \STATE $\mbox{Return TRUE}$
   \ELSE
   \STATE $\mbox{Return FALSE}$
   \ENDIF
\end{algorithmic}
\end{algorithm}


\section{Proof of Lemma~\ref{l:comptime}} \label{ap:comptime}
As we discussed in Section~\ref{s:comp}, in each iteration, the worst computational cost is $O(p^2)$.
Consequently, the total cost is $O(p^3)$.

\section{Proof of Lemma~\ref{l:statprop}} \label{ap:statprop}


Let the Cholesky factors of the sample and the population covariance matrices be
$S = \hat L \hat L^{'}$ and $\Sigma = L L^{'}$.

The from from \citet{olkin1985} the joint distribution of the elements of $\hat L$ is

\begin{equation}
    p(\hat L) = 2^p c \Big( \prod_{i = 1}^p L_{ii}^{-n}\Big) \Big( \prod_{i = 1}^p \hat L_{ii}^{n -p - 1}\Big)
    \Big( \prod_{i = 1}^p \hat L_{ii}^{p-i+1}\Big) \exp \Big( -\frac{1}{2} \mbox{tr} L^{-1} \hat L \hat L^{'} L^{-'}\Big),
\end{equation}

where $c^{-1} = 2^{np/2} \pi^{p(p-1)/4} \prod_{i = 1}^p \Gamma \Big(\frac{n-i+1}{2} \Big)$.

Let $U = L^{-1} \hat L$, then for $i>j$, $U_{ij} \sim N(0,1)$ and $U_{ii} \sim \chi^1_{n-i+1}$.
As a result 
\begin{equation} \label{eq:hatij}
    \hat L_{ij} = L_{ij} U_{jj} + \sum_{l = j+1}^i L_{il} U_{lj}
\end{equation}
\begin{equation} \label{eq:expij}
        E[\hat L_{ij}] =  L_{ij} E[U_{jj}]
\end{equation}
and 
\begin{equation} \label{eq:expii}
    E[\hat L_{ii}] = L_{ii} E[U_{ii}]
\end{equation}
where $E[U_{ii}] = \frac{\sqrt{2} \Gamma[(n-i+2)/2]}{\Gamma[(n-i+1)/2]}$

Consider the random variable $V_{ii} = U_{ii} / \sqrt{n-i+1}$. Since $U_{ii} = \sqrt{U_{ii}^2} = \|Z_1, \dots, Z_{n-i+1}\|_2/\sqrt{n-i+1}$, where $Z_i \sim N(0,1)$ and the Euclidean norm is a 1-Lipschitz function, from \citet[Theorem 2.26]{wainwright2019}
$$P(V_{ii} - E[V_{ii}] \geq t) \leq \exp \Big( - \frac{(n-i+1)t^2}{2}\Big)$$
From Jensen's inequality 
$$E[V_{ii}] \leq \sqrt{E[V_{ii}^2]} = \Big ( \frac{1}{n-i+1} \sum_{k=1}^{n-i+1}E[Z_{k}^2]\Big)^{1/2} = 1$$
Thus,
\begin{equation} \label{eq:vii}
    P(V_{ii} - 1 \geq t) \leq P(V_{ii} - E[V_{ii}] \geq t) \leq \exp   \Big( - \frac{(n-i+1)t^2}{2}\Big)
\end{equation}
and as $n \rightarrow \infty$, $V_ii$ goes to 1 with high probability.
Now,
$$E[\tilde L_{ij}] = L_{ij}E[V_{ij}]$$
and  the result follows from (\ref{eq:vii}).

\section{Proof of Theorem~\ref{t:signcons}} \label{a:cscssign}

The proof of Theorem~\ref{t:signcons} relies on the following assumptions.

\begin{itemize}
\item{A1} \textit{Marginal sub-Gaussian assumption:} The sample matrix $X \in \mathcal{R}^{n \times p}$ has $n$ independent 
rows with each row drawn from the distribution of a zero-mean random vector $X = (X_1, \dots, X_p)^t$ with covariance $\Sigma$ and sub-Gaussian marginals; i.e.,
\[E[\mbox{exp}(tX_j/ \sqrt{\Sigma_{jj}})] \leq \mbox{exp}(Ct^2) \]
for all $j=1,\dots,p, \, t \leq 0$ and for some constant $C > 0$.
\item{A2} \textit{Sparsity Assumption:} The true Cholesky factor $L \in \mathcal{R}^{p \times p}$ is the lower triangular matrix with 
positive diagonal elements and support $\mathcal{S}(L) =\{(i,j), i \neq j| L_{ij} \neq 0\} $. We denote by $s = |S|$ cardinality of the set $S$.
\item{A3} \textit{Bounded eigenvalues:} There exist a constant $\kappa$ such that
\[0 < \kappa^{-1} \leq \lambda_{min}(L) \leq \lambda_{max}(L) \leq \kappa \]
\item{A4} The minimum edge strength:
\begin{equation}
    \ell_{\min}:= \min_{1 \leq j < i \leq p}|L_{ij}| > c_1 \sqrt{\frac{s\log p}{n}}
\end{equation}
\end{itemize}

The following lemma, which proof is for a general penalty functions $\rho(\cdot, \lambda)$
that include $\ell_1$ penalty as in (\ref{eq:objfunc}) and satisfy the conditions below, will be useful in the proof of Theorem~\ref{t:signcons}.

\begin{itemize}
\item The function $\rho(\cdot, \lambda)$ satisfies $\rho(0,\lambda) = 0$ and is symmetric around zero.
\item On the non negative real line, the function  $\rho(\cdot, \lambda)$  is nondecreasing.
\item For $t > 0$, the function $t \rightarrow \rho(\cdot, \lambda) / t$ is nonincreasing in $t$.
\item The function $\rho(\cdot, \lambda)$ is differentiable for all $t \neq 0$ and subdifferentiable at $t = 0$, with $\mbox{lim}_{t \rightarrow 0^+} \rho'(t, \lambda) = \lambda C$.
\item There exists $\mu > 0$ such that $\rho_{\mu}(t, \lambda) = \rho(t, \lambda) + \frac{\mu}{2}t^2$ is convex.
\end{itemize}

Recall that a matrix $\hat L \in \mathcal{L}_p$ is a stationary point for (\ref{eq:objfunc}) if it satisfies \citep{bertsekas2015}
\begin{equation} \label{eq:stat}
\langle\nabla \mathcal{L}_n(\hat L) + \nabla \rho(\hat L, \lambda), L - \hat L \rangle \geq 0,\; \mbox{for}\, L \in \mathcal{L}_p,
\end{equation}
where $\mathcal{L}_n(L) = \mbox{tr}(SLL^t) - 2\log|L|$ and $\nabla \rho(\cdot,\cdot)$ is the subgradient.

\begin{lemma} \label{l:stprop}
Under Assumptions A1-A3, with tuning parameter $\lambda$ of scale $\sqrt{\frac{\log p}{n}}$, and $\frac{3}{4 \gamma} < (\kappa + 1)^{-2}$, 
the scaling $(s+p)\log p = o(n)$ is sufficient for any stationary point $\hat L$ of the (\ref{eq:objfunc}) to satisfy the following estimation bounds:
\[
\begin{aligned}
\|\hat L - L\|_F &= \mathcal{O}_p \Big (\sqrt{\frac{(s + p)\log p}{n}} \Big ) \\
\|\hat L_{\mbox{off}} - L_{\mbox{off}}\|_F &= \mathcal{O}_p \Big (\sqrt{\frac{s \log p}{n}} \Big ), \\
\end{aligned}
\]
where $L_{\mbox{off}}$ refers to all the off-diagonal entries of a matrix $L$.
\end{lemma}
The proof is provided in Section~\ref{l:stprop}.

Now, we are ready to provide the proof of Theorem~\ref{t:signcons}.
From Assumption 2, for $(i,j), i\neq j \not \in \mathcal{S}(L)$, we have $L_{ij} = 0$ and 
from Lemma~\ref{l:stprop} there exist $c_2 > 0$ such that 
$$|\hat L_{ij}| < c_2 \frac{ s \log p}{n}$$
Then by the definition of thresholded CSCS (\ref{eq:thresh}), it follows that
$\hat L^{\tau} = 0$.

For $(i,j) \in \mathcal{S}(L)$, from the Assumption 4  
$$
L_{ij} > c_1 \sqrt{\frac{s\log p}{n}}
$$
and from Lemma~\ref{l:stprop} for $i \neq j$ 
$$|\hat L_{ij} - L_{ij}| < c_2 \sqrt{\frac{s \log p}{n}}.$$
Since, we assumed $c_1 > 2 c_2$, it follow that
$|\hat L_{ij}| > c_2 \sqrt{\frac{s \log p}{n}}$.
Then by the definition of thresholded CSCS (\ref{eq:thresh}), it follows that
$\hat L^{\tau} \neq 0$ and the result follows.

\subsection{Proof of Lemma~\ref{l:stprop}}

We start by showing that $\mathcal{L}_n$ satisfies RSC conditions. Recall that the differentiable function $\mathcal{L}_n: \mathcal{R}^{p \times p} \rightarrow \mathcal{R}$ satisfies RSC condition if:
\[
\langle \nabla \mathcal{L}_n(L + \Delta) - \nabla \mathcal{L}_n(L), \Delta \rangle \geq
\]
\begin{numcases}{\geq}
 	\alpha_1 \| \Delta \|^2_F - \tau_1 \frac{\log p}{n} \| \Delta \|^2_1, & $\forall \| \Delta \|_F \leq 1$  \label{eq:rsc1}\\
        \alpha_2 \| \Delta \|_F - \tau_2 \sqrt{\frac{\log p}{n}} \| \Delta \|_2, & $\forall \| \Delta \|_F \geq 1$  \label{eq:rsc2}
\end{numcases}
 where the $\alpha_j$'s are strictly positive constants and the $\tau_j$'s are nonnegative constants. From \citet[Lemma 4]{loh2015}, 
 under conditions of Lemma~\ref{l:stprop}, if (\ref{eq:rsc1}) holds then (\ref{eq:rsc2}) holds. 
 Thus, we concentrate only on showing that  (\ref{eq:rsc1}) holds for $\| \Delta \|_F \leq 1$.
 Recall that
 \begin{equation} \label{e:ln}
 \mathcal{L}_n(L) = \mbox{tr}(SL^tL) - 2\log|L|
\end{equation}

\begin{lemma} \label{l:rsc}
 The cost function (\ref{e:ln}) satisfies RSC condition with $\alpha_1 = (\kappa + 1)^{-2}$ and $\tau_1 = 0$; i.e.,
 \begin{equation} \label{eq:rscp}
  \langle \nabla \mathcal{L}_n(L + \Delta) - \nabla \mathcal{L}_n(L), \Delta \rangle \geq  (\kappa + 1)^{-2} \| \Delta \|^2_F, \: \forall \| \Delta \|_F \leq 1
 \end{equation}
\end{lemma}
The proof is provided in Section~\ref{a:rsc}.

From the penalty conditions listed above, $\rho_{\mu}(L,\lambda) = \rho(L, \lambda) + \frac{\mu}{2} \|L\|^2_F$ is convex. Thus,
\[
\begin{aligned}
\rho_{\mu}(L,\lambda) - \rho_{\mu}(\hat L,\lambda) &\geq \langle \nabla \rho_{\mu}(\hat L,\lambda), L - \hat L \rangle \\
&=\langle \nabla \rho(\hat L,\lambda) + \mu \hat L, L - \hat L \rangle, 
\end{aligned}
\]
which implies that
\begin{equation} \label{eq:rho}
 \langle \nabla \rho( \hat L,\lambda), L - \hat L \rangle \leq \rho( L,\lambda) - \rho( \hat L,\lambda) + \frac{\mu}{2} \|\hat L - L\|^2_F
\end{equation}

From stationarity condition (\ref{eq:stat})
\[\langle \nabla \mathcal{L}_n (\hat L), L - \hat L \rangle \geq - \langle \nabla \rho(\hat L, \lambda), L - \hat L \rangle\]
and combining above result with (\ref{eq:rscp})
\[ \begin{aligned}
(1 + \kappa)^{-2}\|\Delta\|^2_F &\leq \langle \mathcal{L}_n(\hat L), \Delta \rangle - \langle \nabla \mathcal{L}_n(L), \Delta \rangle \\
& \leq \langle \nabla \rho(\hat L, \lambda), L - \hat L \rangle - \langle \nabla \mathcal{L}_n(L), \Delta \rangle  \\
& \leq \rho(L, \lambda) - \rho(\hat L, \lambda) + \frac{\mu}{2} \|\hat L - L\|^2_F  \\
&- \langle \nabla \mathcal{L}_n(L), \Delta \rangle
\end{aligned}
\]
After rearrangement and H{\''o}lder inequality
\[
\begin{aligned}
\Big((1 + \kappa)^{-2}) - \frac{\mu}{2} \Big) \|\Delta\|^2_F & \leq \rho(L, \lambda) - \rho(\hat L, \lambda)\\ & + \|\nabla \mathcal{L}_n(L)\|_{\infty} \|\Delta\|_1
\end{aligned}
\]
From \citet[Lemma 4]{loh2015}
\[\lambda \|\Delta\|_1 \leq \rho(\Delta,\lambda) + \frac{\mu}{2}\|\Delta\|^2_F\]
and from \citet[Lemma 15]{yu2017} under the assumed scaling of $\lambda$
\[\|\nabla \mathcal{L}_n(L)\|_{\infty} \leq \frac{\lambda}{2}\]
with probability going to 1. Combining above two results and using subadditive property; i.e., $\rho(\Delta,\lambda) \leq \rho(L, \lambda) + \rho(\hat L, \lambda)$:
\[
\begin{aligned}
\Big( (1 + \kappa)^{-2} - \frac{\mu}{2} \Big ) \|\Delta\|^2_F &\leq \rho(L, \lambda) - \rho (\hat L, \lambda) + \frac{\lambda}{2} \|\hat \Delta\|_1\\
& \leq \rho(L, \lambda) - \rho (\hat L, \lambda) \\
&+ \frac{\rho (\Delta, \lambda)}{2} + \frac{\mu}{4} \|\Delta\|^2_F \\
& \leq \rho(L, \lambda) - \rho(\hat L, \lambda)\\
&+ \frac{\rho (L, \lambda) + \rho(\hat L, \lambda)}{2} + \frac{\mu}{4} \|\Delta\|^2_F
\end{aligned}
 \]
After rearranging and using $3/4\mu \leq (1+ \kappa)^{-2} $
\begin{equation} \label{eq:cnd1}
0 \leq \Big( (1 + \kappa)^{-2} - \frac{3}{4}\mu \Big) \|\Delta\|^2_F \leq 3 \rho(L, \lambda) - \rho(\hat L, \lambda)
\end{equation}
From (\ref{eq:cnd1}) and \citet[Lemma 5]{loh2015} follows
\[ \rho(L, \lambda) - \rho(L, \lambda) \leq 2\lambda \|\Delta_S\| - \lambda \|\Delta_{S^c}\| \Rightarrow \|\Delta_{S^c}\|_1 \leq 3 \|\Delta_S\|_1\]
Thus,
\[
\begin{aligned}
\Big(2 (1 + \kappa)^{-2}  - \frac{3}{2}\mu \Big) \|\Delta\|^2_F  &  \leq \lambda \|\Delta_S\|_2 - \lambda \|\Delta_{S^c}\|_1 \\
& \leq \lambda \|\Delta_S\|_1 \leq \lambda \sqrt{p + s} \|\Delta\|_F,
\end{aligned}
\]
from which we conclude that
\begin{equation} \label{eq:res1}
\|\Delta\|_F \leq \frac{6\lambda \sqrt{p + s}}{4(1 + \kappa)^{-2} - 3 \mu},
\end{equation}
and the result follows from the chosen scaling of $\lambda$.

For the precision matrix bound, from page 45 of \cite{yu2017} we note that
\[  \hat L^t \hat L - L^tL = (\hat L - L)^t (\hat L - L) + (\hat L - L)^TL + L^t (\hat L - L)\]
and
\[\|L^t (\hat L - L)\|_F \leq |||L|||_2 \|\hat L - L\|_F\]
From submultiplicativity property of matrix norm
\[\|(\hat L - L)^t (\hat L - L)\|_F \leq \|(\hat L - L)\|^2_F \]
Therefore,
\begin{equation} \label{eq:res2}
 \|\hat L^t \hat L - L^tL\|_F \leq (\|\hat L - L\|_F + 2 |||L|||_2)\|\hat L - L\|_F
 \end{equation}

The upper bound for the off-diagonal elements $\hat L_{\mbox{off}}$ easily follows from the
above proof. For example, see \citet[Proposition 1]{wangetal2022}.

\subsection{Proof of Lemma~\ref{l:rsc}} \label{a:rsc}
The following facts will be useful in the proof.

\textbf{Fact 1}

\begin{enumerate}
\item $(K_{pp})^{-1} = K_{pp}$
\item $\lambda_{max}(K_{pp}) = 1$
\item $\mbox{tr}(ABCD) = \mbox{vec}(D^t)(C^t \otimes A)\mbox{vec}(B)$
\item $\lambda_{max}(A \otimes B) = \lambda_{max}(A)\lambda_{max}(B)$
\end{enumerate}
where $K_{pp}$ is the commutation matrix  such that $\mbox{vec}(L) = K_{pp} \mbox{vec}(L^t)$. The proof of the facts can be found in \citet[Section 4]{magnus1986}.

To show the RSC condition, we rely on the directional derivatives (for example see \citet[Section 6.3]{tao2016}). In particular, if we denote by $D_{\Delta}\mathcal{L}_n(L)$ the directional derivative with respect to the direction $\Delta$, then from \citet[Lemma 6.3.5]{tao2016} :
\begin{equation} \label{e:dder1}
\langle \nabla \mathcal{L}_n(L), \Delta \rangle = D_{\Delta}\mathcal{L}_n(L) = 2\mbox{tr}[(SL^t - L^{-1})\Delta]
\end{equation}
Similarly
\begin{equation} \label{e:dder2}
\begin{aligned}
\langle \nabla \mathcal{L}_n(L + \Delta), \Delta \rangle &= D_{\Delta}\mathcal{L}_n(L + \Delta) \\
&= 2\mbox{tr}[(S(L+\Delta)^t - (L+ \Delta)^{-1})\Delta]
\end{aligned}
\end{equation}
From Woodbury identity \cite{Horn2012}
\[(L + \Delta)^{-1} = L^{-1} - L^{-1}\Delta(L + \Delta)^{-1}\]
Plugging back into (\ref{e:dder2}) and after some algebra
\begin{equation} \label{e:dder2.1}
\begin{aligned}
\langle \nabla \mathcal{L}_n(L + \Delta), \Delta \rangle &= 2\mbox{tr}[(S(L+\Delta)^t - (L+ \Delta)^{-1})\Delta] \\
&+ 2\mbox{tr}[S \Delta^t\Delta + L^{-1}\Delta(L + \Delta)^{-1}\Delta]
\end{aligned}
\end{equation}
Thus, from (\ref{e:dder1}) and (\ref{e:dder2.1})

\begin{equation} \label{eq:dder3}
\begin{aligned}
\langle \nabla \mathcal{L}_n(L + \Delta) &- \nabla \mathcal{L}_n(L), \Delta \rangle =\\ &= 2\mbox{tr}[\Delta^tS\Delta + L^{-1}\Delta(L+\Delta)^{-1}\Delta] \\
& \geq \mbox{vec}(\Delta)^t K_{pp}((L + \Delta)^{-t}\otimes L^{-1}) \mbox{vec}(\Delta)\\
& = \mbox{vec}(\Delta)^t [((L + \Delta)^{t}\otimes L)K^{-1}_{pp}]^{-1} \mbox{vec}(\Delta)\\
& \geq \lambda_{min}([((L + \Delta)^{t}\otimes L)K^{-1}_{pp}]^{-1})\|\Delta\|^2_F,
\end{aligned}
\end{equation}
where for the first inequality we used the fact that $S$ is positive semi-definite and the second equality follows from the Fact 1.
Now, since
\begin{equation} \label{eq:preres}
\begin{aligned}
\lambda_{min}([((L + \Delta)^{t}&\otimes L)K^{-1}_{pp}]^{-1}) =\\
&= \lambda^{-1}_{max}[((L + \Delta)^{t}\otimes L)K^{-1}_{pp}] \\
&\geq \lambda^{-1}_{max}(K^{-1}_{pp}) \lambda^{-1}_{max}(L) \lambda^{-1}_{max}(L + \Delta)\\
&\geq (\kappa + 1)^{-2} ,
\end{aligned}
\end{equation}
where the first inequality follows from the submultiplicativity property of the norm and lower-triangularity of the $L$ and $\Delta$. The second inequality follows from the triangular property, the fact that $\|\Delta\|_2 \leq \|\Delta\|_F \leq 1$ and, properties of the $K_{pp}$ stated in Fact 1. After plugging (\ref{eq:preres}) into (\ref{eq:dder3}), the result follows.

\section{Assumption~\ref{a:majorass} and varsortability} \label{a:varsort}
In this section, we show that Assumption~\ref{a:majorass} holds even when varsortability is low.
Consider the following data generation process

\begin{equation} \label{eq:dgp1}
    \begin{aligned}
        X_1 &:= \varepsilon_1\\
        X_2 &:= X_1 + \varepsilon_2\\
        X_3 &:= 0.85X_2 + \varepsilon_3\\
        X_4 &:= 0.79X_2 + \varepsilon_4,
    \end{aligned}
\end{equation}
where $\sigma^2_1 = 4$, $\sigma^2_2 = 2$, $\sigma^2_3 = 1$, and $\sigma^2_{4} = 1.5$. 
That is $x = \{4,2,1,1.5\}$, where $x$ is defined as in (\ref{e:x}).

After generating data from (\ref{eq:dgp1}) with $n = 1000$ and using the function \texttt{varsortability()} 
available from \url{https://github.com/Scriddie/Varsortability/}, it can be shown that, for this case, varsortability is equal to 0.6.
Recall that when varsitability is one then ordering by marginal variance is a valid causal ordering.

We use procedure described in Appendix~\ref{s:quickmajor} to check that Assumption~\ref{a:majorass} holds. 
Table~\ref{t:varcheck} reports the vectors $y_{\rho}$, where for each permutation $\rho \in \mathfrak{S}_p$,
$y_{\rho}$ is defined as in (\ref{e:y}). As can be seen, from Proposition~\ref{p:majcheck}, $x \prec_w y_{\rho}$ is true
for all permutations.

\begin{table}[ht]
\centering
    \begin{tabular}{c|c}
    $\rho$ & $y_{\rho}$ \\
    \hline
$\{1,2,4,3\}$ & $\{4.00 , 2.00 , 1.50 , 1.00\}$ \\ 
 $\{1,4,2,3\}$ & $\{4.00 , 2.75 , 1.09 , 1.00\}$ \\ 
  $\{4, 1, 2,3\}$ & $\{5.24 , 2.10 , 1.09 , 1.00\}$ \\ 
  $\{4,1,3,2\}$ & $\{5.24 , 2.10 , 1.79 , 0.61 \}$ \\ 
  $\{1,4,3,2\}$ & $\{4.00 , 2.75 , 1.79 , 0.61\}$ \\ 
  $\{1,3,4,2\}$ & $\{4.00 , 2.44 , 2.01 , 0.61\}$ \\ 
  $\{1,3,2,4\}$ & $\{4.00 , 2.44 , 1.50 , 0.82\}$ \\ 
  $\{3,1,2,4\}$ & $\{5.33 , 1.83 , 1.50 , 0.82\}$ \\ 
  $\{3,1,4,2\}$ & $\{5.33 , 2.01 , 1.83 , 0.61\}$ \\ 
  $\{3,4,2,1\}$ & $\{5.33 , 2.20 , 1.67 , 0.61\}$\\ 
  $\{4,3,1,2\}$ & $\{5.24 , 2.24 , 1.67 , 0.61\}$ \\ 
  $\{4,3,2,1\}$ & $\{5.24 , 2.24 , 1.33 , 0.77\}$ \\ 
  $\{3,4,2,1\}$ & $\{5.33 , 2.20 , 1.33 , 0.77\}$ \\ 
  $\{3,2,4,1\}$ & $\{5.33 , 1.50 , 1.33 , 1.12\}$ \\ 
  $\{3,2,1,4\}$ & $\{5.33 , 1.50 , 1.33 , 1.12\}$ \\ 
  $\{2,3,1,4\}$ & $\{6.00 , 1.50 , 1.33 , 1.00\}$ \\ 
  $\{2,3,4,1\}$ & $\{6.00 , 1.50 , 1.33 , 1.00\}$ \\ 
  $\{2,4,3,1\}$ & $\{6.00 , 1.50 , 1.33 , 1.00\}$ \\ 
  $\{4,2,3,1\}$ & $\{5.24 , 1.72 , 1.33 , 1.00\}$ \\ 
  $\{4,2,1,3\}$ & $\{5.24 , 1.72 , 1.33 , 1.00\}$ \\ 
  $\{2,4,1,3\}$ & $\{6.00 , 1.50 , 1.33 , 1.00\}$ \\ 
  $\{2,1,4,3\}$ & $\{6.00 , 1.50 , 1.33 , 1.00\}$ \\ 
  $\{2,1,3,4\}$ & $\{6.00 , 1.50 , 1.33 , 1.00\}$ \\ 
   \hline
\end{tabular}
    \caption{Estimated vectors $y_{\rho}$ for the data generation process (\ref{eq:dgp1}).}
    \label{t:varcheck}
\end{table}

\section{Bayesian Networks} \label{s:bn}

We start by introducing the following graphical concepts. If the graph $\mathcal{G}$ contains a directed edge from the node $k \rightarrow j$, then $k$ is a parent of its child $j$. We write $\Pi^{\mathcal{G}}_{j}$ for the set of all parents of a node $j$. If there exists a directed path $k \rightarrow \dots \rightarrow j$, then $k$ is an ancestor of its descendant $j$. 
A \textit{Bayesian Network} is a directed acyclic graph $\mathcal{G}$ whose nodes represent random variables $X_1, \dots, X_p$. Then $\mathcal{G}$ encodes a set of conditional independencies and conditional probability distributions for each variable. The DAG $\mathcal{G} = (V,E)$ is characterized by the node set $V = \{1, \dots, p\}$ and the edge set $E = \{(i,j): i \in \Pi^{\mathcal{G}}_j\} \subset V \times V$. It is well-known that for a BN, the joint distribution factorizes as:

\begin{equation} \label{eq:markov}
P(X_1,\dots,X_p) = \prod_{j=1}^{p}P(X_j| \Pi^{\mathcal{G}}_{j})
\end{equation}

 \begin{figure}[ht!]
\centering
\includegraphics[width= 9cm, height = 5cm]{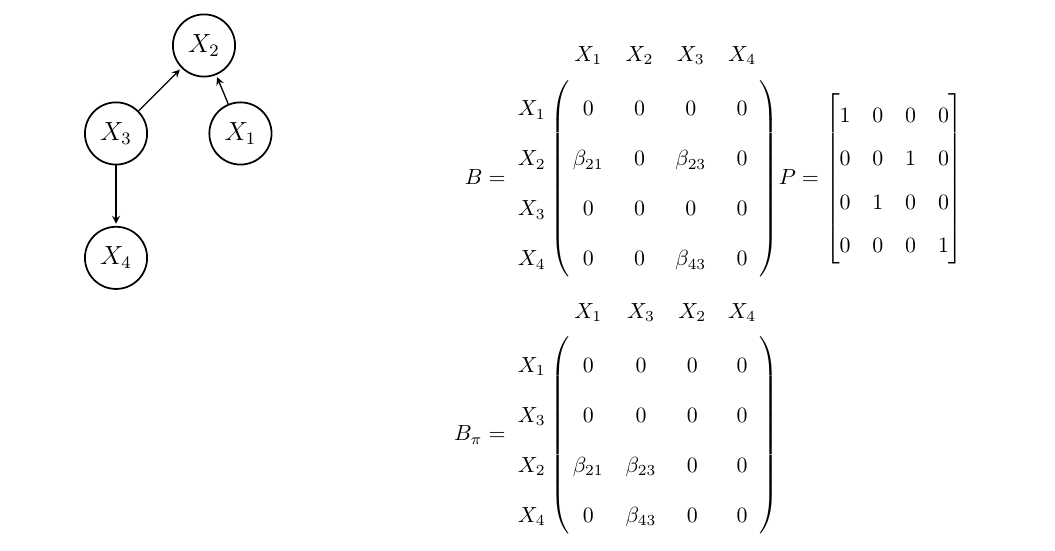}
\caption{Illustration of DAG $\mathcal{G}$, corresponding coefficient matrix $B$, permutation matrix $P$, and permuted strictly lower triangular matrix $B_{\pi}$.}
\label{fig:exdag}
\end{figure}

\section{Additional Results}

\subsection{Bank Connedtedness analysis for LINGAM} \label{a:adresult}
\begin{figure}[ht!]
      \centering
          \includegraphics[width=0.6\textwidth]{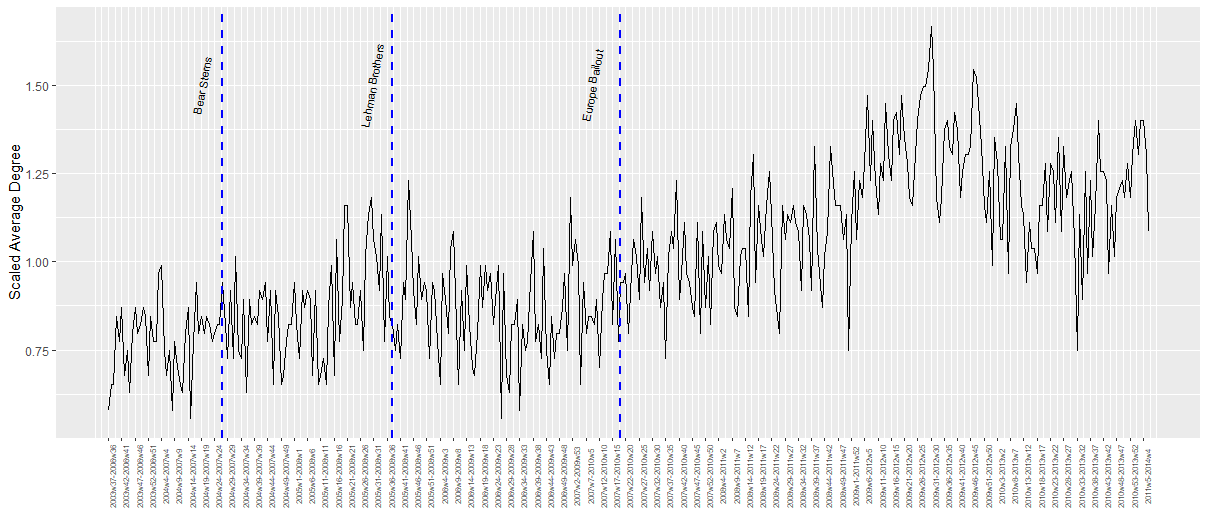}
         \caption{Evolution of average degree of bank connectedness scaled by their
             historic average (over 2003- 2014) using LINGAM.}
         \label{fig:avg_degree_macho_LINGAM}
 \end{figure}

Figure~\ref{fig:avg_degree_macho_LINGAM} illustrates average degrees for the network estimated using LINGAM.
As can be seen, the plot is noisy and no significant pattern can be discerned.

\begin{figure}[ht!]
    \centering
    \includegraphics[width = 0.8\textwidth]{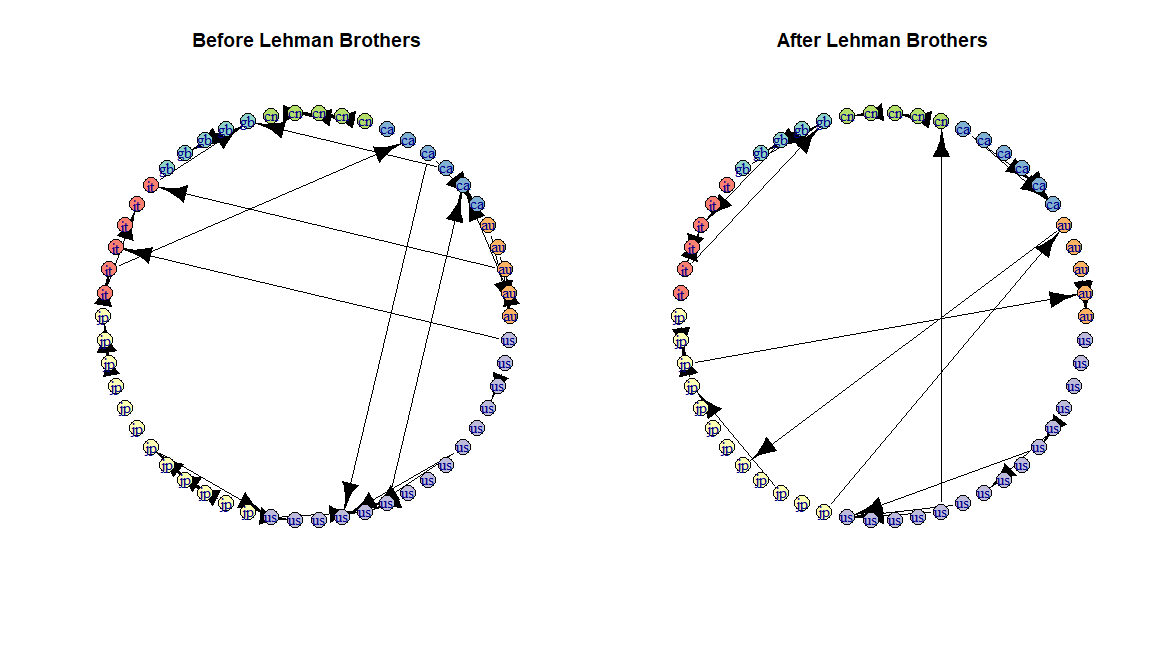}
    \caption{The DAG estimated from LINGAM does not show significant changes between before and 
    after Lehman Brothers failure.}
    \label{fig:daglingam}
\end{figure}

\subsection{Multivariate Linear Simulation Results} \label{a:addsim}

\begin{figure}[ht!]
     \centering
     \begin{subfigure}[b]{0.8\textwidth}
         \centering
         \includegraphics[width=\textwidth]{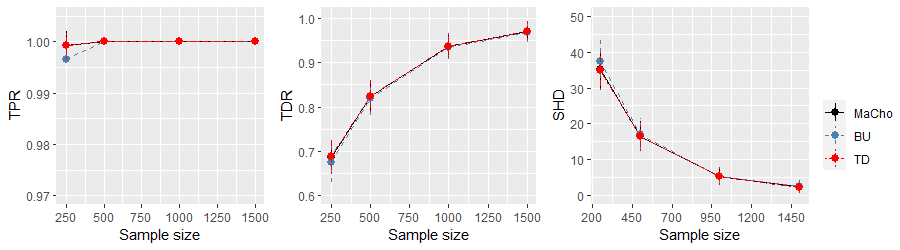}
         \caption{Case 1}
         \label{fig:case1p20}
     \end{subfigure}
     \vfill
     \begin{subfigure}[b]{0.8\textwidth}
         \centering
         \includegraphics[width=\textwidth]{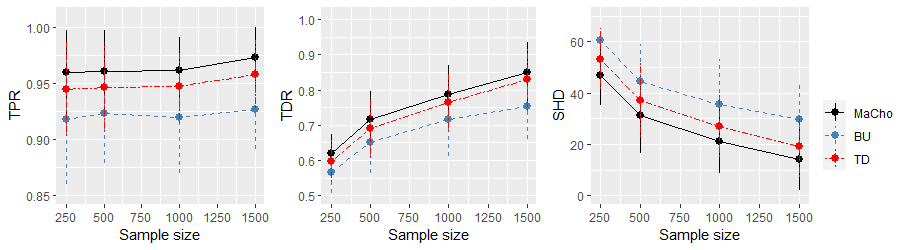}
         \caption{Case 2}
         \label{fig:case2p20}
     \end{subfigure}
     \vfill
        \begin{subfigure}[b]{0.8\textwidth}
         \centering
         \includegraphics[width=\textwidth]{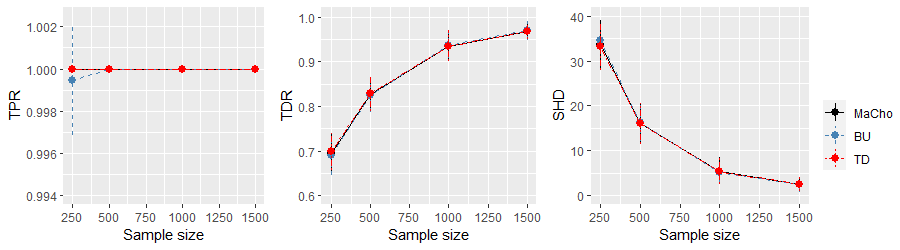}
         \caption{Case 3}
         \label{fig:case3p20}
     \end{subfigure}
         \caption{Comparison of the proposed algorithm MaCho, BU, and TD algorithms in terms 
         of average TPR, TDR, and SHD for recovering linear SEM with different error variances and $p = 20$.}
        \label{fig:p20result_LING}
\end{figure}

\begin{figure}[ht!]
     \centering
     \begin{subfigure}[b]{0.8\textwidth}
         \centering
         \includegraphics[width=\textwidth]{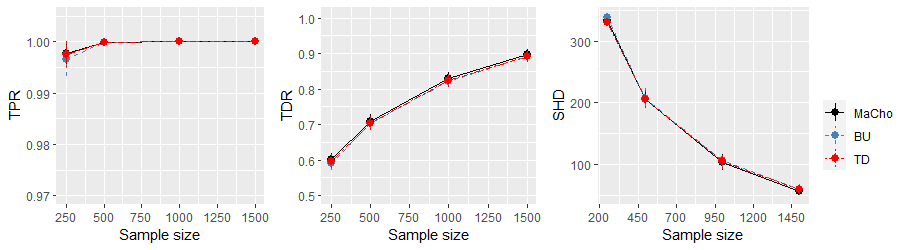}
         \caption{Case 1}
         \label{fig:case1p50}
     \end{subfigure}
     \vfill
     \begin{subfigure}[b]{0.8\textwidth}
         \centering
         \includegraphics[width=\textwidth]{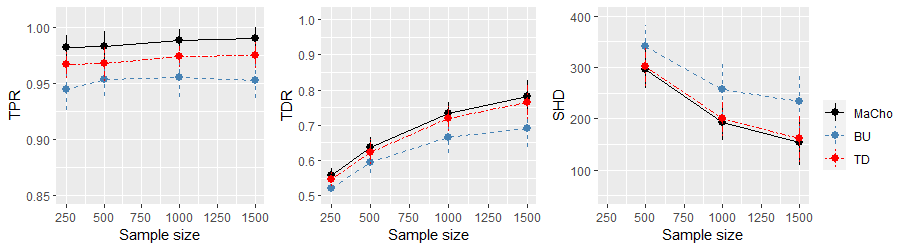}
         \caption{Case 2}
         \label{fig:case2p50}
     \end{subfigure}
     \vfill
        \begin{subfigure}[b]{0.8\textwidth}
         \centering
         \includegraphics[width=\textwidth]{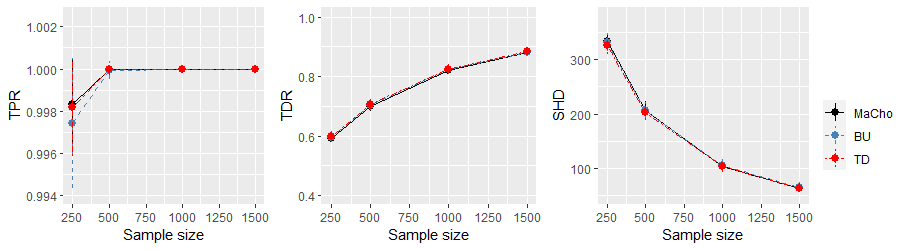}
         \caption{Case 3}
         \label{fig:case3p50}
     \end{subfigure}
         \caption{Comparison of the proposed algorithm MaCho, BU, and TD algorithms in terms 
         of average TPR, TDR, and SHD for recovering linear SEM with different error variances and $p = 50$.}
        \label{fig:p50result_LING}
\end{figure}

Figures~\ref{fig:p20result_LING} and \ref{fig:p50result_LING}   report the results for $p=20$ and $p=50$ without LINGAM algorithm.

\end{document}